\documentclass[twocolumn,trackchanges]{aastex63}

\usepackage{longtable}
\usepackage{xstring}
\usepackage{colordvi}
\usepackage{pmboxdraw}

\received{April 20, 2020}
\revised{May 10, 2020}
\accepted{\today}
\submitjournal{ApJS}

\shorttitle{Spectroscopic follow-up}
\shortauthors{Boutsia et al.}
\graphicspath{{./}{figures/}}

\newcommand{\nqso}{N_{\rm QSO}}
\newcommand{\sv}{\sigma_v}
\newcommand{\ti}{t_{\rm int}}
\newcommand{\zqso}{z_{\rm QSO}}

\newcommand{\NqubricsI}[1]{
\IfEqCase{#1}{
{candidates}{1476}
{flagA}{327}
{QSO}{313}
{QSOzGT25}{203}}
{surveycovered}{22\%}
{successrate}{62\%}
}

\newcommand{\Nqubrics}[1]{
\IfEqCase{#1}{
{mainsample}{1014875}
{Selected}{1412}
{Candidates}{818}
{SelObserved}{266}
{KnownSel}{594}
{Completeness}{95\%}
{SuccessRate}{68\%}
{tot}{569}          
{dupl}{58}          {dupl_F1}{10.2\%}             
{dupl_n}{8}         {dupl_n_F2}{13.8\%}           {dupl_n_F1}{1.4\%}            
{dupl_d}{29}        {dupl_d_F2}{50.0\%}           {dupl_d_F1}{5.1\%}            
{dupl_g}{21}        {dupl_g_F2}{36.2\%}           {dupl_g_F1}{3.7\%}            
{unique}{511}       {unique_F1}{89.8\%}           
{flagA}{432}        {flagA_F2}{84.5\%}            {flagA_F1}{75.9\%}            
{A_QSO}{390}        {A_QSO_F3}{90.3\%}            {A_QSO_F2}{76.3\%}            {A_QSO_F1}{68.5\%}            
{A_QSO_zl25}{166}   {A_QSO_zl25_F4}{42.6\%}       {A_QSO_zl25_F3}{38.4\%}       {A_QSO_zl25_F2}{32.5\%}       {A_QSO_zl25_F1}{29.2\%}       
{A_QSO_zg25}{224}   {A_QSO_zg25_F4}{57.4\%}       {A_QSO_zg25_F3}{51.9\%}       {A_QSO_zg25_F2}{43.8\%}       {A_QSO_zg25_F1}{39.4\%}       
{A_QSO_zg35}{54}    {A_QSO_zg35_F4}{13.8\%}       {A_QSO_zg35_F3}{12.5\%}       {A_QSO_zg35_F2}{10.6\%}       {A_QSO_zg35_F1}{9.5\%}        
{A_QSO_zg40}{15}    {A_QSO_zg40_F4}{3.8\%}        {A_QSO_zg40_F3}{3.5\%}        {A_QSO_zg40_F2}{2.9\%}        {A_QSO_zg40_F1}{2.6\%}        
{A_GAL}{11}         {A_GAL_F3}{2.5\%}             {A_GAL_F2}{2.2\%}             {A_GAL_F1}{1.9\%}             
{A_STAR}{31}        {A_STAR_F3}{7.2\%}            {A_STAR_F2}{6.1\%}            {A_STAR_F1}{5.4\%}            
{flagB}{79}         {flagB_F2}{15.5\%}            {flagB_F1}{13.9\%}            
{B_QSO}{18}         {B_QSO_F3}{22.8\%}            {B_QSO_F2}{3.5\%}             {B_QSO_F1}{3.2\%}             
{B_QSO_zl25}{13}    {B_QSO_zl25_F4}{72.2\%}       {B_QSO_zl25_F3}{16.5\%}       {B_QSO_zl25_F2}{2.5\%}        {B_QSO_zl25_F1}{2.3\%}        
{B_QSO_zg25}{5}     {B_QSO_zg25_F4}{27.8\%}       {B_QSO_zg25_F3}{6.3\%}        {B_QSO_zg25_F2}{1.0\%}        {B_QSO_zg25_F1}{0.9\%}        
{B_QSO_zg35}{1}     {B_QSO_zg35_F4}{5.6\%}        {B_QSO_zg35_F3}{1.3\%}        {B_QSO_zg35_F2}{0.2\%}        {B_QSO_zg35_F1}{0.2\%}        
{B_QSO_zg40}{1}     {B_QSO_zg40_F4}{5.6\%}        {B_QSO_zg40_F3}{1.3\%}        {B_QSO_zg40_F2}{0.2\%}        {B_QSO_zg40_F1}{0.2\%}        
{B_MOSTRO}{31}      {B_MOSTRO_F3}{39.2\%}         {B_MOSTRO_F2}{6.1\%}          {B_MOSTRO_F1}{5.4\%}          
{B_STAR}{7}         {B_STAR_F3}{8.9\%}            {B_STAR_F2}{1.4\%}            {B_STAR_F1}{1.2\%}            
{B_???}{18}         {B_???_F3}{22.8\%}            {B_???_F2}{3.5\%}             {B_???_F1}{3.2\%}             
{B_GAL}{5}          {B_GAL_F3}{6.3\%}             {B_GAL_F2}{1.0\%}             {B_GAL_F1}{0.9\%}             
}[\PackageError{Nqubrics}{Undefined input: #1}{}]}

\newcommand{\CCA}{\texttt{CCA}}

\begin{document}

\title{The spectroscopic follow-up of the QUBRICS bright quasar survey}

\correspondingauthor{Konstantina Boutsia}
\email{kboutsia@carnegiescience.edu}

\author{Konstantina Boutsia}
\affil{Las Campanas Observatory, Carnegie Observatories, 
Colina El Pino, Casilla 601, La Serena, Chile\\}

\author{Andrea Grazian}
\affil{INAF--Osservatorio Astronomico di Padova, 
Vicolo dell'Osservatorio 5, I-35122, Padova, Italy\\}

\author{Giorgio Calderone}
\affil{INAF--Osservatorio Astronomico di Trieste, 
Via G.B. Tiepolo, 11, I-34143 Trieste, Italy \\}

\author{Stefano Cristiani}
\affil{INAF--Osservatorio Astronomico di Trieste, 
Via G.B. Tiepolo, 11, I-34143 Trieste, Italy \\}
\affiliation{INFN-National Institute for Nuclear Physics,  
via Valerio 2, I-34127 Trieste \\}
\affil{IFPU--Institute for Fundamental Physics of the Universe, via Beirut 2, I-34151 Trieste, Italy}

\author{Guido Cupani}
\affil{INAF--Osservatorio Astronomico di Trieste, 
Via G.B. Tiepolo, 11, I-34143 Trieste, Italy \\}

\author{Francesco Guarneri}
\affil{INAF--Osservatorio Astronomico di Trieste, 
Via G.B. Tiepolo, 11, I-34143 Trieste, Italy \\}
\affil{Dipartimento di Fisica, Sezione di Astronomia, Università di Trieste, 
via G.B. Tiepolo 11, I-34131, Trieste, Italy}

\author{Fabio Fontanot}
\affil{INAF--Osservatorio Astronomico di Trieste, 
Via G.B. Tiepolo, 11, I-34143 Trieste, Italy \\}

\author{Ricardo Amorin}
\affil{Instituto de Investigaci\'on Multidisciplinar en Ciencia y Tecnolog\'ia, Universidad de La Serena, 
Raul Bitr\'an 1305, La Serena, Chile}
\affil{Departamento de F\'isica y Astronom\'ia, Universidad de La Serena, 
Av. Juan Cisternas 1200 Norte, La Serena, Chile}

\author{Valentina D'Odorico}
\affil{INAF--Osservatorio Astronomico di Trieste, 
Via G.B. Tiepolo, 11, I-34143 Trieste, Italy \\}
\affil{Scuola Normale Superiore, 
P.zza dei Cavalieri, I-56126 Pisa, Italy\\}
\affil{IFPU--Institute for Fundamental Physics of the Universe, via Beirut 2, I-34151 Trieste, Italy}

\author{Emanuele Giallongo}
\affil{INAF--Osservatorio Astronomico di Roma, Via Frascati 33, I-00078, Monte Porzio Catone, Italy}

\author{Mara Salvato}
\affil{Max-Planck-Institut für extraterrestrische Physik Giessenbachstrasse 1, Garching D-85748, Germany}

\author{Alessandro Omizzolo}
\affil{Specola Vaticana, Vatican Observatory, 
00122 Vatican City State\\}
\affil{INAF--Osservatorio Astronomico di Padova, 
Vicolo dell'Osservatorio 5, I-35122, Padova, Italy\\}

\author{Michael Romano}
\affil{Dipartimento di Fisica e Astronomia, Universit\`a di Padova,
Vicolo dell'Osservatorio 3, I-35122, Padova, Italy}
\affil{INAF--Osservatorio Astronomico di Padova, 
Vicolo dell'Osservatorio 5, I-35122, Padova, Italy\\}

\author{Nicola Menci}
\affil{INAF--Osservatorio Astronomico di Roma, Via Frascati 33, I-00078, Monte Porzio Catone, Italy}

\begin{abstract}
We present the results of the spectroscopic follow up of the QUBRICS\footnote{Acronym for {\it ``QUasars as 
BRIght beacons for Cosmology in the Southern hemisphere''}, \citep{calderone19}.} survey. 
The selection method is based on a machine learning approach applied to photometric catalogs, 
covering an area of $\sim$ 12,400 deg$^2$ in the Southern Hemisphere.
The spectroscopic observations started in 2018 and identified 55 new,
high-redshift ($z \geq 2.5$), bright ($i \leq 18$) QSOs, with the catalog published in late 2019.
Here we report the current status of the survey,
bringing the total number of bright QSOs at $z \geq 2.5$ identified by QUBRICS to\Nqubrics{A_QSO_zg25}.
The success rate of the QUBRICS selection method, in its most recent training,
is estimated to be\Nqubrics{SuccessRate}.
The predominant contaminant turns out to be lower-z QSOs at $z<2.5$.
This survey provides a unique sample of bright QSOs at high-z
available for a number of cosmological investigations.
In particular, carrying out the redshift drift measurements
(Sandage Test) in the Southern Hemisphere,
using the HIRES spectrograph at the 39m ELT, appears to be possible with less than 2500 hours of 
observations spread over 30 targets in 25 years.

\end{abstract}


\section{Introduction} \label{sec:intro}

Luminous quasars are the brightest non-transient
cosmic beacons in the Universe. The hunt for such bright sources,
especially at high redshift, is of paramount importance for a number
of extragalactic studies, ranging from the number density of bright
quasars at high-z
\citep{Sch19a}, the theoretical modelling of the early phases of galaxy formation and 
co-evolution with their central SMBHs \citep[e.g.][]{Vali16,Fon2020}, the study and 
characterisation of their (gas) accretion properties
of the super massive black hole (SMBH) population \citep{Wu15, Wolf18} to the inference 
on cosmological parameters from time delays of
strongly lensed QSOs \citep{Bonvin17} and the properties of the dark matter
by microlensing statistics in bright quasars \citep{Webster91,Bate07}.

Absorption signatures in the spectra of bright high-z QSOs are one of
the most powerful and invaluable tools for studying intergalactic
environments, as emerged from the recent Astro2020 Decadal Survey
\citep[e.g.][]{Beck19}. Among the fundamental questions that can be tackled thanks to the 
study of QSO absorption lines we recall: the measurement of primordial Deuterium 
abundance \citep[e.g.][]{Cooke18}, the temperature evolution of the Cosmic Microwave 
Background (CMB), the free-streaming of warm dark matter \citep[e.g.][]{Irsic17}, the 
variation of the fundamental constants of nature, e.g. the fine structure constant or 
the proton-to-electron mass ratio \citep[see][]{Leite16}, the missing baryon problem \citep[e.g.][]{Werk14}, 
the production and diffusion of metals in the IGM \citep[e.g.][]{Dod16}, the Lyman continuum 
escape fraction of high-z QSOs \citep[e.g.][]{Cristiani16,Grazia18}, the mean free path of 
ionizing photons \citep[e.g.][]{Prochaska09,Worseck14,Romano19},
the reionization epochs of hydrogen and helium, and the sources responsible of these 
transition phases in cosmic history.

An appealing application of the detailed study of the Lyman forest in
Cosmology is the so-called Sandage test \citep{Sandage62}, which can give
fundamental constraints to General Relativity. The
detection of the tiny drift due to cosmic expansion in the cosmological redshifts of many
absorption lines in the spectra of bright QSOs will allow to measure
directly cosmological parameters (e.g. $\Omega_M$, $\Omega_\Lambda$,
$H_0$) at $2<z<5$ without the need of any local ladders or
intermediate distance indicators. This revolution will be possible
only with the brightest QSOs observed by the most powerful and stable
high resolution spectrographs which will be available in the future at
20-40m telescopes \citep{Liske08}.

However, finding the brightest quasars at high-z is not a trivial process.
The advent of the SDSS survey \citep[e.g.][]{fan2000} represents a quantum
leap in this respect, at least in the Northern
hemisphere. At present, the SDSS has delivered more than $10^6$
spectroscopically confirmed QSOs at $0<z<6.5$, with a large fraction
at absolute magnitudes $M_{1450}\le -26$. Recent studies, however,
point out that, at very bright magnitudes, SDSS can suffer from
incompleteness due to color selection (see also \cite{Fontanot07}). 
For example, \citet{Sch19a} find 407 new bright
QSOs at $2.8<z<5.0$ in the ELQ survey, showing that the SDSS
completeness is $\sim 60\%$ at bright magnitudes ($i\le 18$). As a
consequence, the hunt for bright quasars, especially at high-z and in
the Northern hemisphere, could be biased towards lower numbers due to
the adoption of efficient but low completeness selections.

The situation is even more dramatic in the Southern hemisphere, due to
the lack of wide multi-wavelength surveys at $\delta\le 0^\circ$ in
the past. Comparing QSO surface densities, it is statistically
evident that high-z objects of bright apparent magnitudes
must be present also in the Southern hemisphere: of the 22 known QSOs
with $z>3$ and $V<17$, only 5 are at $\delta < 0^\circ$, and all the 3
QSOs with $V<16$ are in the North \citep{Veron10}.

In \citet{calderone19} (hereafter Paper I), we presented the first results of
a survey of $z \geq 2.5$ QSOs at bright i-band magnitudes ($i\le 18.0$) in the Southern
hemisphere, taking advantage of the recent availability of new multiwavelength public databases.
The combination of state-of-the-art
databases with innovative techniques for the selection of the best
candidates results in an efficient selection, with a success rate in finding high-z QSOs 
larger than $50 \%$ and a completeness in excess of 90\%. 
In the first spectroscopic runs we already identified the intrinsically most luminous QSO 
at that time, with $z\ge 3.8$ at $\delta < 0^\circ$, QSO J2157-3602, which has been afterwards 
independently confirmed by \citet{Sch19b}. In this paper we will present a new z$>$4 QSO with 
apparent magnitude of $i=16.886$ (see Section 4) that would be the new record holder in the 
Southern Hemisphere. In total we present 168 new bright QSOs at $z \geq 2.5$, thus quickly 
completing the identification of the high-z sample of Paper I.

The structure of this paper is the following: in Section 2 we
summarize the selection method of our bright QSO candidate sample, in
Section 3 and 4 we describe the massive observational campaign carried
out at medium size telescopes and the spectroscopic identification of
our QSO candidates. The properties of the newly identified high-z
QSOs are discussed in Section 5, while Section 6 is devoted to the
description of the Golden QSO sample for the Sandage test. Discussions
and Conclusions are summarized in Section 7. Unless otherwise stated, apparent
magnitudes are in the AB photometric system.

\section{Selection method} \label{sec:selection}
In this section we will describe the QUasars as BRIght beacons for Cosmology in the Southern 
hemisphere (QUBRICS) survey, first introduced in Paper I. We refer the reader to this work for 
more details of the method, while here we just recall its main characteristics. 

To identify new, high-$z$ QSOs candidates in the remaining sample, a classification algorithm 
has been applied based on the Canonical Correlation Analysis, \citep[CCA,][]{CCA}. Our 
candidate list has been drawn from a multiwavelength catalogue. We used the public 
databases of: i) Skymapper (DR1.1, \cite{Wolf18}); ii) Gaia (DR2, \cite{Gaia18}); 
iii) 2MASS \citep{Ref2MASS} and iv) the WISE survey \citep{Wri10} to build the 
initial {\it main sample} (\Nqubrics{mainsample} sources), covering $\sim$~12400 
deg${^2}$, with $i$-band magnitudes in the range $14 < m_{i} \leq 18$. 
The sources with secure object-type identification have been used as a training set 
and the recipe has been applied to the remaining sources in order to predict a likely 
classification. By using the parallax and proper motion estimates provided by Gaia, $\sim$~83\% of 
the sources were classified as {\it bona fide} stars.  Matching the remaining sources against 
the following catalogs: SDSS DR14Q \citep{Paris18}, Veron-Cetty QSO 13th edition \citep{Veron10}, 
and 2dFGRS \citep{Colless01}, led to reliably identify an object-type classification 
to 4666 QSOs and 3665 galaxies (in the following, the {\it original training set}). 
To discriminate against low-$z$ ($<2.5$) QSOs, we use the CCA as a regression algorithm to 
the objects classified as QSOs, in the previous step, to obtain an estimate of their redshift. 
In Paper I we tested this methodology with a resulting sample of 1476 candidates without 
spectroscopic confirmation, and our spectroscopic follow-up identified 54 QSOs with $z \geq 2.5$. 

\begin{figure*}[]
\epsscale{1}
\plotone{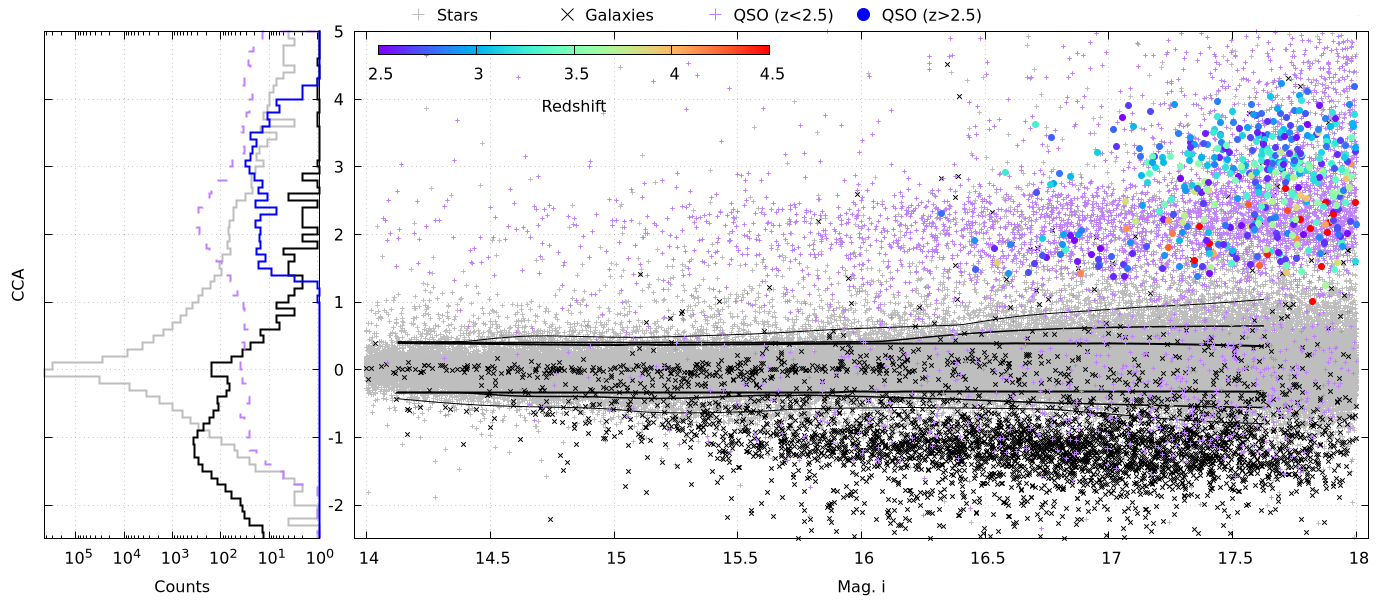}\\
\plotone{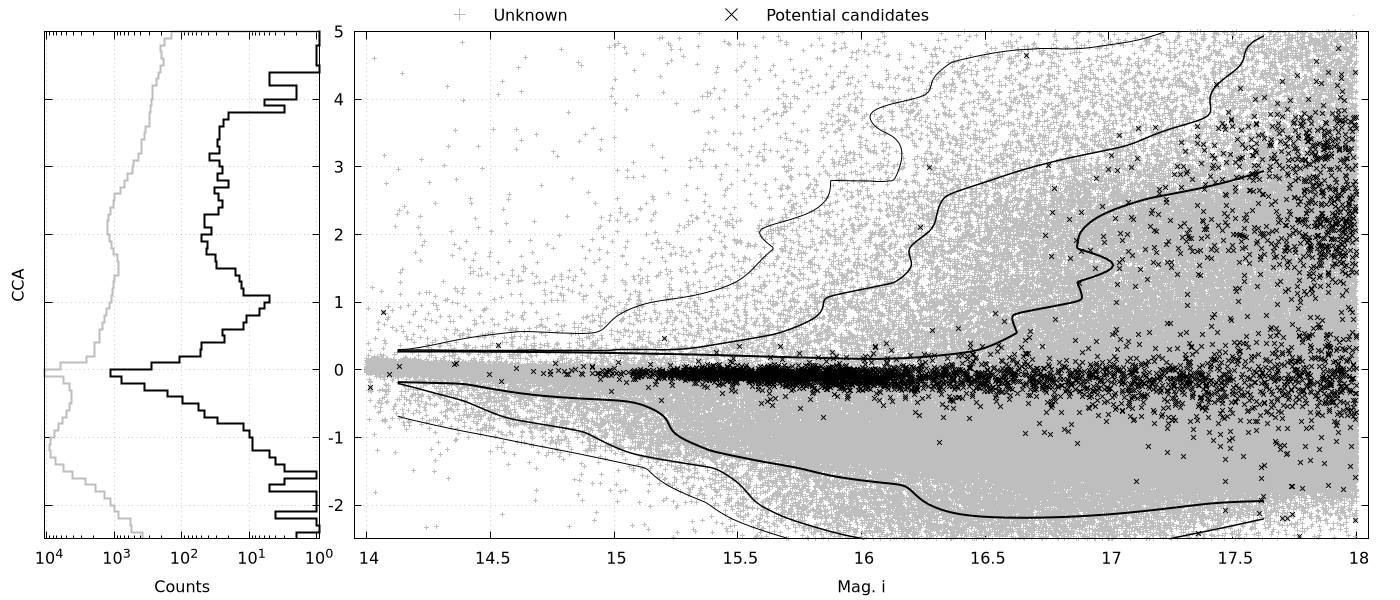}\\
\plotone{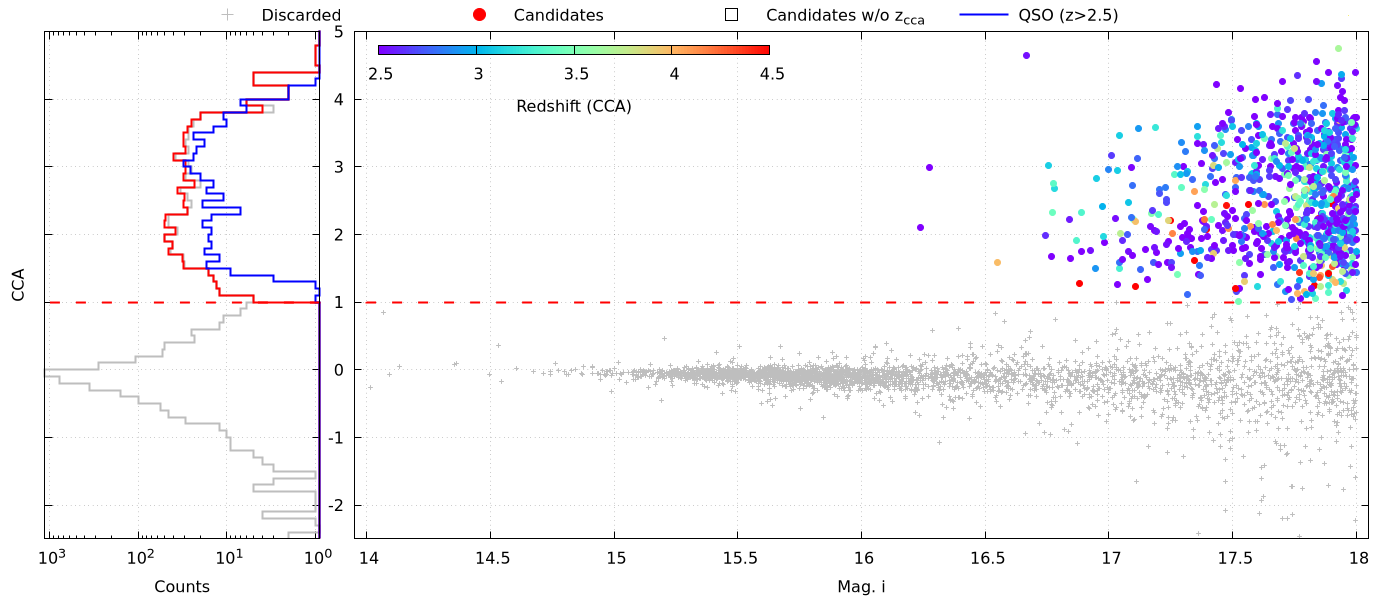}\\
\caption{The \CCA{}--$i$ plane for the various subsamples considered in this work.
For the statistical meaning of the CCA parameter shown in the y-axis we refer the reader to Paper 
I. Upper panel: sources in the {\it main} sample for which a reliable type identification is 
available. Stars are identified by gray ``+'' symbols, inactive galaxies by black cross 
symbols, low-$z$ ($<2.5$) QSOs with purple ``+'' symbols, high-$z$ ($\geq 2.5$) QSOs with 
filled circles. The redshifts for the spectroscopically confirmed QSOs with $z \geq 2.5$ are 
shown with the color code shown in the colorbox in the upper left corner. The inset on the left 
shows the histogram of the \CCA{} coordinate for the stars (gray), galaxies (black), low-$z$ 
QSOs (purple) and known high-$z$ QSOs (blue). Middle panel: sources in the {\it main sample} 
without an object type identification (gray symbols). The same sources after excluding extended 
and low-$z$ objects are highlighted in black, and represents potential high-$z$ QSO candidates. 
Lower panel: the {\it final} sample of high-$z$ QSO candidates, with their redshift $z_{\rm cca}$ 
as in Paper I. The red dashed line represents the cut in the CCA selection. Candidates/observed 
sources are represented with open and filled circles respectively.  The gray histogram represents 
the potential candidates, while the red and blue ones represent the QSO candidates and the confirmed QSO samples.}
\label{Fig:cca1}
\end{figure*}

\begin{figure}[]
  \plotone{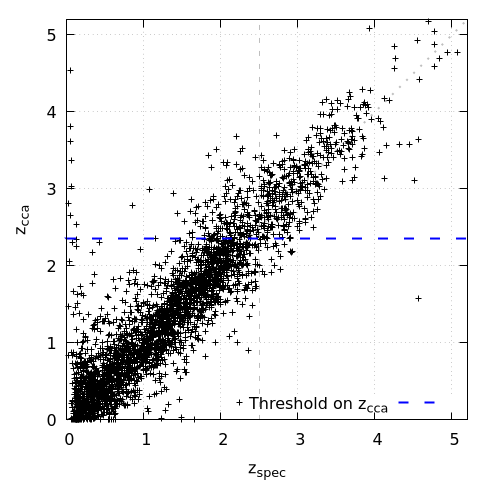}\\
  \plotone{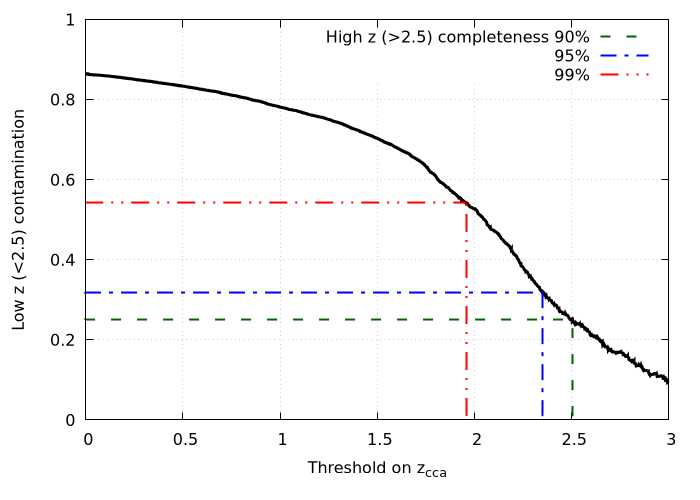}\\
  \caption{Upper panel: The $z_{\rm cca}$--$z_{\rm spec}$ correlation (scatter: $\sim$~0.37). 
  The horizontal blue line is the threshold on $z_{\rm cca}$ = 2.26, corresponding to a 
  completeness of 95\% and a contamination of 38\%.  Lower panel: contamination and completeness 
  as a function of the threshold on $z_{\rm cca}$.}
\label{Fig:zcca}
\end{figure}

As the main focus of our work is the identification of the largest possible number of bright 
high-z QSO in the Southern Emisphere, we aim at the highest possible success rate of observing 
runs. Therefore, after each run we update the training set of the algorithm by including the 
new identifications: this results in an evolving list of QSO candidates. For the statistical 
purposes of this paper, the list of candidates has been frozen to its current form after the 
last observing run in Feb.~2020. At this point the list included\Nqubrics{Selected} sources 
in total, with\Nqubrics{KnownSel} secure spectroscopic identifications and\Nqubrics{Candidates} 
candidates yet to be observed. 
Thanks to the revised training set it has been possible to reduce the number of candidates 
yet to be observed by $\sim$~25\% with respect to the original list of Paper I. Fig.~\ref{Fig:cca1} 
and \ref{Fig:zcca} illustrate the process of classification and redshift estimate, respectively, 
in the present work. They update the corresponding plots of Paper I.

It should be noted that the selection criteria adopted here and in Paper I could be biased against 
lensed sources. To reduce the contamination we have adopted relative stringent criteria about the 
positional coincidence in the various photometric catalogs. While this is not affecting "normal" QSOs, may have 
the subtle effect of removing lensed sources, forming extended structures like Einstein rings or crosses. 
Moreover, due to the choice of the fundamental photometric bands (e.g. GAIA), our selections is 
probably biased against sources at $z\gtrsim 4.5$. 

\section{Spectroscopy} \label{sec:spectroscopy}

The QUBRICS pilot campaign has been presented in Paper I. Observations have been obtained using various 
instruments at Las Campanas Observatory (LCO), the TNG telescope, and the ESO-NTT telescope at La Silla. 
Between August 2019 and February 2020, we have been awarded more nights at these facilities, in order to 
expand our spectroscopic survey. We discuss here the details of these observations.

\subsection{WFCCD at duPont}
We have been awarded a total of 12 nights with the Wide Field CCD camera at duPont. We used the same 
configuration as in the pilot study, namely, the blue grism, with the 1.5" long slit, that covers a 
wavelength range between 3700 - 8000 {\AA} giving a 2 {\AA}/pixel dispersion. In total we have observed 
100 candidates, but only 76 of them have a robust spectroscopic classification (flag=A). Of the securely 
classified sources, 63 (83\%) are QSOs, and 36 of those (57\%) are in the desired redshift range (z$ \geq $2.5).

\subsection{LDSS-3 at Clay}
A total of 44 candidates have been observed with LDSS-3 at the Clay telescope. Observations were 
obtained in several nights during bright time and variable weather conditions. We have used the 
VPH-all grism with the 1" - central slit and no blocking filter, covering a wavelength range between 
4000 - 10000 {\AA} with a low resolution of R$\sim$800. Exposure times were ranging between 800 - 1800sec, 
depending on the candidate magnitude. Out of the observed candidates, 40 have a high quality redshift flag (flag A) 
and have been securely classified, most of them as QSOs (28 out of 40, 70\%) but only 12 (43\%) have a redshift above 2.5. 

\subsection{EFOSC2 at NTT}

In September 2019 we were awarded three more nights (PI. A. Grazian, proposal 0103.A-0746)
at NTT, to use the EFOSC2 instrument. In order to complete the survey, we obtained 4 additional nights 
at NTT in January 2020 (PI. A Grazian, proposal 0104.A-0754). Again, we used grism \#13 (wavelength 
range $\lambda\sim 3700-9300$ {\AA}), with typical exposure times ranging between 300 and 600 seconds. 
We observed 217 candidates, obtaining robust identification for 187. Out of these, 161 (86\%) were 
classified as QSOs and 122 have a redshift of $z \geq 2.5$ (76\%). This was one of the most efficient runs in this period.

The outcome of this new spectroscopic campaign shows that, although our criteria are very efficient at 
identifying QSOs (84\% of robust classifications are QSOs), only 55\% are in the targeted redshift range. 
Anyhow, as discussed in Section 2, by ingesting the new QSOs to the selection algorithm, the efficiency of 
the selection for future spectroscopy is expected to be as high as 75\% in the global sample. 

\subsection{Data Reduction}

WFCCD data were reduced using standard IRAF tasks. After subtracting bias and dividing with flat, 
individual exposures were combined to the final image. We then used the task {\it apall} to extract 
the spectra and standard and {\it sensfunc} to calibrate in flux. The EFOSC2 and LDSS-3 spectra have 
been reduced with a custom pipeline using MIDAS scripts. The standard pre-reduction (i.e. bias subtraction 
and flat field normalization) has been adopted, and the wavelength calibration with helium, neon, and argon 
lamp has been obtained. We also check our wavelength solution on the emission night sky.
An rms of 0.5 {\AA} has been obtained from the wavelength calibration process.
For WFCCD, EFOSC2 and LDSS-3, we made sure to obtain at least one spectro-photometric standard star per night. 
Since conditions have not always been photometric, the derived flux calibration is relative. There has been 
no further attempt to interpolate flux to known broad band magnitudes for absolute flux calibration. Further 
details can be found in Paper I.
Improved data reduction of 3 objects observed in the runs of Paper I 
allowed us to upgrade their redshift flag to A.
They are listed in Tab.\ref{tab:NewSpec} with an appropriate identification in footnote.

\subsection{Redshifts from other surveys}
\label{ref:other_surv}
\begin{figure*}[hbt!]
\epsscale{1}
\plotone{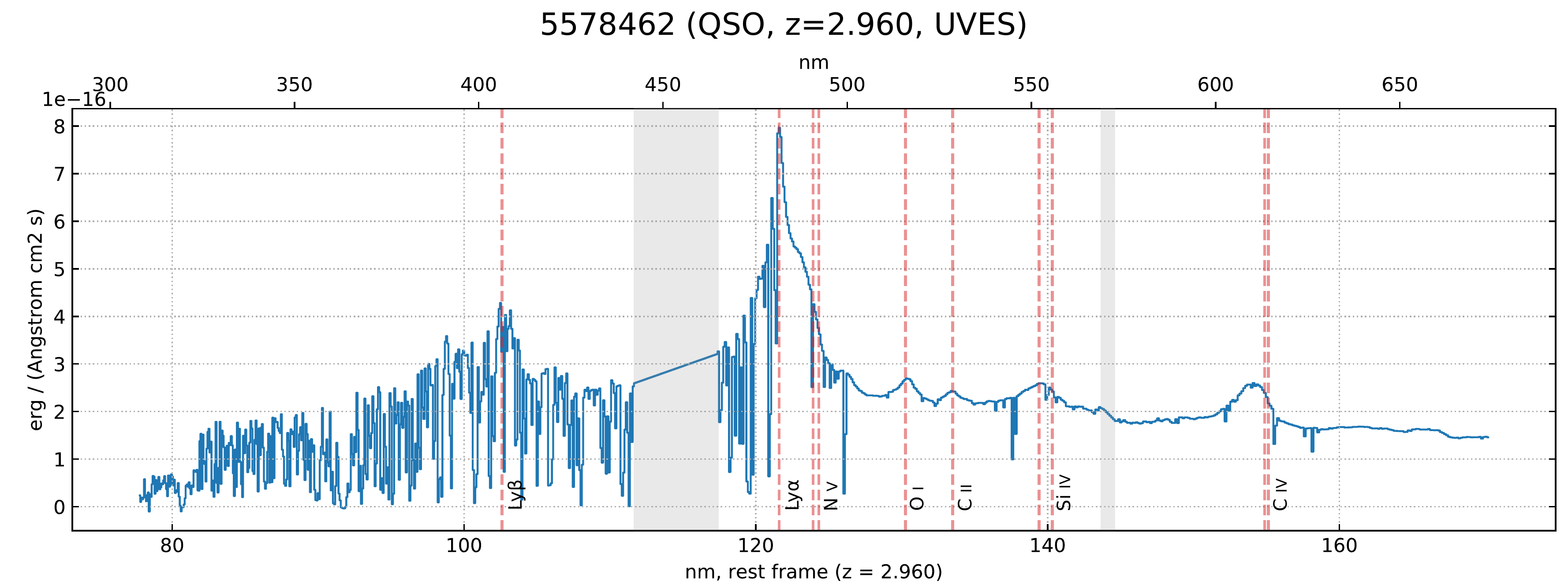}\\
\caption{Composite UVES spectrum of a candidate at z=2.960, with some AGN emission lines highlighted. The grey bands 
are gaps in the UVES wavelength coverage, depending on the adopted instrument setup. 
}
\label{Fig:uves}
\end{figure*}

We have also searched the databases in the literature for additional reliable spectra of our candidates, finding 24 
QSOs (6 with $z\geq 2.5$) and 1 AGN.
They are listed in Tab.~\ref{tab:AddSpectra}.
In particular we found that 
downloading and analyzing spectra from the 6dF survey \citep{6dF09}
we could assign a reliable redshift to 22 of our candidates.
The candidate with SkyMapper ID 5578462 has spectral data of good quality in the ESO archive
(see Fig.~\ref{Fig:uves}) that we have reprocessed.
For the candidate with ID 10779504 we found a reliable redshift determination in the OzDES survey \citep{DES_17}. 

\begin{deluxetable*}{rcccccc}
\tablecaption{Additional identifications from the literature.
}
\label{tab:AddSpectra}
\tablewidth{700pt}
\tabletypesize{\scriptsize}
\tablehead{
\colhead{Skymapper} & \colhead{R.A. (J2000)} & \colhead{DEC. (J2000)} & 
\colhead{$m_i$} & \colhead{$z_{\rm spec}$} &  \colhead{Class} & 
\colhead{Survey} \\
\colhead{ID} & \colhead{} & \colhead{} &
\colhead{(mag)} & \colhead{} & \colhead{} & 
\colhead{} \\
} 
\startdata
  8414380 & 00:43:23.42 & -00:15:52.5 & 17.781 & 1.442 & QSO & 6dF \\
  8133493 & 00:44:48.56 & -13:39:13.0 & 17.664 & 2.253 & QSO & 6dF \\
  7885395 & 00:58:20.54 & -19:04:03.2 & 17.983 & 1.861 & QSO & 6dF \\
  9164524 & 02:03:03.20 & -07:06:04.7 & 17.839 & 1.458 & QSO & 6dF \\
  8826213 & 02:08:41.52 & -19:44:03.7 & 17.943 & 2.137 & QSO & 6dF \\
  8970643 & 02:29:45.41 & -18:17:11.5 & 17.246 & 1.817 & QSO & 6dF \\
315606464 & 03:06:13.56 & -57:51:05.3 & 17.925 & 2.066 & QSO & 6dF \\
318330107 & 03:14:59.84 & -45:45:28.6 & 17.967 & 2.198 & QSO & 6dF \\
 10372687 & 03:26:13.41 & -31:00:52.2 & 17.869 & 2.105 & QSO & 6dF \\
 10779504 & 03:39:53.34 & -27:00:53.4 & 17.988 & 2.410 & QSO & DES \\
 10460382 & 03:40:42.83 & -34:00:44.3 & 17.800 & 1.944 & QSO & 6dF \\
 10812487 & 03:47:14.85 & -24:38:08.7 & 17.427 & 3.130 & QSO & 6dF \\
 11616376 & 04:22:43.67 & -29:25:29.9 & 17.767 & 2.529 & QSO & 6dF \\
 12053071 & 05:05:55.83 & -29:30:38.5 & 17.865 & 1.070 & QSO & 6dF \\
 56866269 & 10:18:21.75 & -21:40:07.8 & 17.547 & 2.441 & QSO & 6dF \\
 57772960 & 11:18:10.70 & -17:51:59.3 & 15.119 & 0.215 & AGN & 6dF \\
 63944442 & 12:12:18.99 & -25:47:26.1 & 17.620 & 2.532 & QSO & 6dF \\
 65530384 & 12:28:48.22 & -01:04:14.7 & 17.039 & 2.642 & QSO & 6dF \\
 64459508 & 12:31:32.96 & -14:36:30.9 & 16.888 & 2.418 & QSO & 6dF \\
 65808873 & 13:25:09.61 & -08:04:48.2 & 17.786 & 2.359 & QSO & 6dF \\
308760834 & 22:21:10.25 & -44:31:57.3 & 17.501 & 2.071 & QSO & 6dF \\
  1459087 & 23:06:37.40 & -36:49:26.0 & 17.962 & 2.650 & QSO & 6dF \\
308911502 & 23:21:22.33 & -50:28:17.5 & 17.701 & 2.297 & QSO & 6dF \\
  5578462 & 23:21:28.67 & -10:51:22.3 & 17.769 & 2.958 & QSO & ESO \\
\enddata
\end{deluxetable*}

%
%
\begin{figure}[]
  \plotone{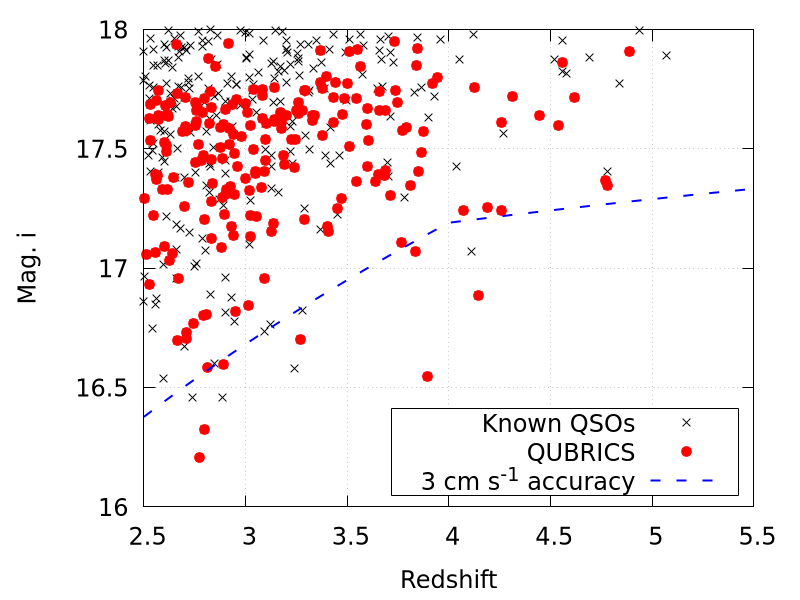}
  \caption{The redshift-$i$ magnitude plane of the QSOs in the area of the present survey. Black crosses: QSOs 
  known before the present observations; Red filled circles: new spectroscopic redshifts obtained in QUBRICS. 
  The dashed blue line shows, in the context of the Sandage Test, the locus of an accuracy of 3 cm/s in radial 
  velocity, reachable allocating 2500 h of observations at the ELT to a single QSO
(see Sect.~\ref{sec:SandageT} for details).
}
\label{Fig:i_vs_z}
\end{figure}

\section{Results of the spectroscopy}
\label{sec:spectro-results}

After 26 observing runs (including those reported in Paper I) we have collected the spectra of \Nqubrics{unique} 
sources, and\Nqubrics{flagA} of them have a secure object-type classification. In total we identified\Nqubrics{A_QSO_zg25} 
new bright QSOs at $z \geq 2.5$ (of which\Nqubrics{A_QSO_zg40} at $z>4$),\Nqubrics{A_QSO_zl25} 
QSOs/AGN at $z<2.5$,\Nqubrics{A_GAL} inactive galaxies and\Nqubrics{A_STAR} stars. 
Among the observed sources, a small fraction has uncertain classification and/or redshift estimate. 
These sources have been assigned a flagB for a variety of reasons. More precisely 32\% have flat spectra 
without clear emission features that could be either stars or galaxies. Another 68\% have a tentative QSOs 
classification, but it is not possible to robustly estimate the redshift since there is only one emission 
line or they show broad absorption features that are difficult to interpret. Such sources will be the subject 
of additional observations in different wavelength ranges aiming to discover additional features that could 
facilitate classification. An example of such follow-up can be seen in Section 4.3.

The list of the 303 new sources, to be added to those reported in Paper I, and their basic properties
are shown in Tab.~\ref{tab:NewSpec}.

The success rate, defined as the fraction of high-$z$ QSOs among all the sources with a secure classification, 
as selected by our algorithm (\S\ref{sec:selection}), is relatively high (412 / 726 = 57\%) for the old Paper I 
list and 405 / 594 $\sim$~68\% 
for the present work. The main contaminant (in the case of the present work
the only one) is represented by low-$z$ ($z<2.5$) QSOs/AGNs.
If we consider only the\Nqubrics{Selected} selected sources as described in Sect.~\ref{sec:selection}, 266 have 
been observed and 199 turned out to be $z \geq 2.5$ QSOs, corresponding to a success rate of 75\%. 

In the present work, the success rate could be biased towards higher values, due to the self-learning approach 
described in Section \ref{sec:selection}. If we consider the effective success rate, i.e. the global 
ratio of high-$z$ QSOs identifications carried out by us over all spectroscopic observations done so far, 
including empirical attempts on improbable candidates, we still obtain a remarkable 52\%. 

All machine learning algorithms are biased by the features of the initial input sample. In this survey, the 
initial training set has been created based on all known spectroscopically confirmed QSOs from the literature, 
selected with a wide range of methods. Thus, bias towards specific types of sources should be minimal. However,
in order to better assess the properties of our (evolving) selections and achieve a better training, we have 
carried out observations also of some objects not complying with our initial specifications, for example with 
no GAIA counterpart within the established radius or not point-like according to our criteria or fainter than 
18mag in the i band ($i>18$), thus not included in the Main Sample.
These objects are also listed in Tab.~\ref{tab:NewSpec}
which contains two columns to show whether the object is a candidate according to the criteria of Paper I or/and 
selected according to the present work. Sources not included in the Main Sample are indicated with asterisks. 

The total number of QSOs with $z \geq 2.5$ (and $i \leq 18$) in the QUBRICS {\it main sample} is 428, with 202 
discovered by our survey and 226 sources derived from the literature. The machine learning algorithm applied in 
Paper I was able to identify 412 of the 428 $z \geq 2.5$ known QSOs (96\%). Following the self-learning 
approach, i.e. by re-ingesting all new identifications in the training set, this completeness indicator becomes $405/428=95\%$.

\subsection{QSOs at z $\geq$ 2.5}
Tab.~\ref{tab:NewSpec} lists 168 new bright QSOs with spectroscopic redshift z$ \geq $2.5, and considering the sources 
already published in PaperI, the total number of new bright QSOs discovered by our survey amounts to \Nqubrics{A_QSO_zg25}.

\begin{figure*}
  \plotone{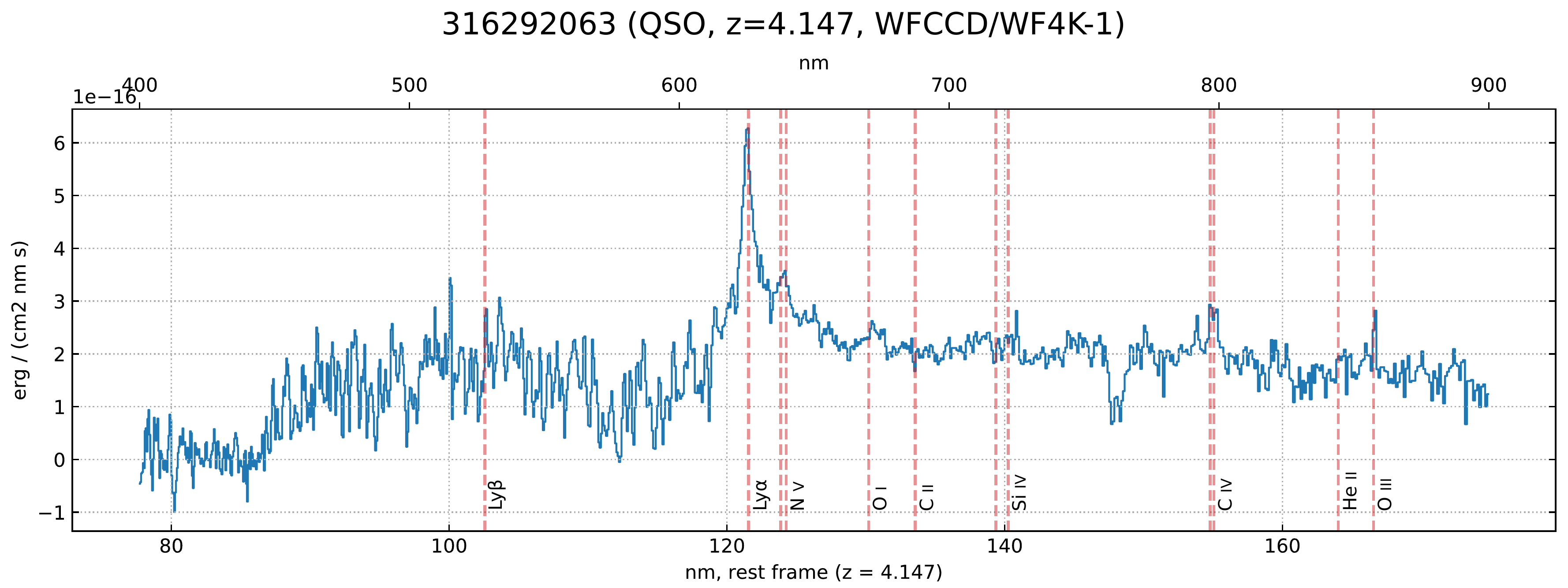}
  \caption{Spectrum, resampled at 250 km/s, of the brightest z$\sim$4 QSO in the sample, with some AGN emission lines }
\label{Fig:spec316}
\end{figure*}

Fig.\ref{Fig:i_vs_z} shows the updated redshift-magnitude diagram for $z \geq 2.5$ quasars in the area of the QUBRICS 
survey. Red points indicate sources published in Paper I or here. 
Out of the 10 new QSOs at $z \geq 4.0$ the brightest is 316292063  with a z=4.147 and $m_{i}$=16.88,
shown in Fig.~\ref{Fig:spec316}. As seen in the figure, this QSO has rather narrow lines and this could perhaps 
indicate a type2 QSO or a lensed source. Additional observations have already been requested and its spectrum will 
be discussed in detail once our follow-up is completed. Our list includes 5 sources that were already published 
by \citet{Wolf19} and were independently selected and classified by our survey. 
\subsection{QSOs at z $<2.5$}
Out of the 121 sources with flag A that are not high-z QSOs, there are 25 with redshift $z \leq 0.5$ and 78 QSOs at $0.5<z<2.5$. 
Thus 85\% of the sources (103 out of 121) are active galaxies, but not in our desired redshift range. 
These relatively bright objects may still turn out to be useful for studies of the evolution of the metal content of the IGM,
and of the Lyman forest at low redshift with space observations.

\subsection{Notes on BAL QSOs}
\begin{figure*}[hbt!]
\epsscale{1}
\plotone{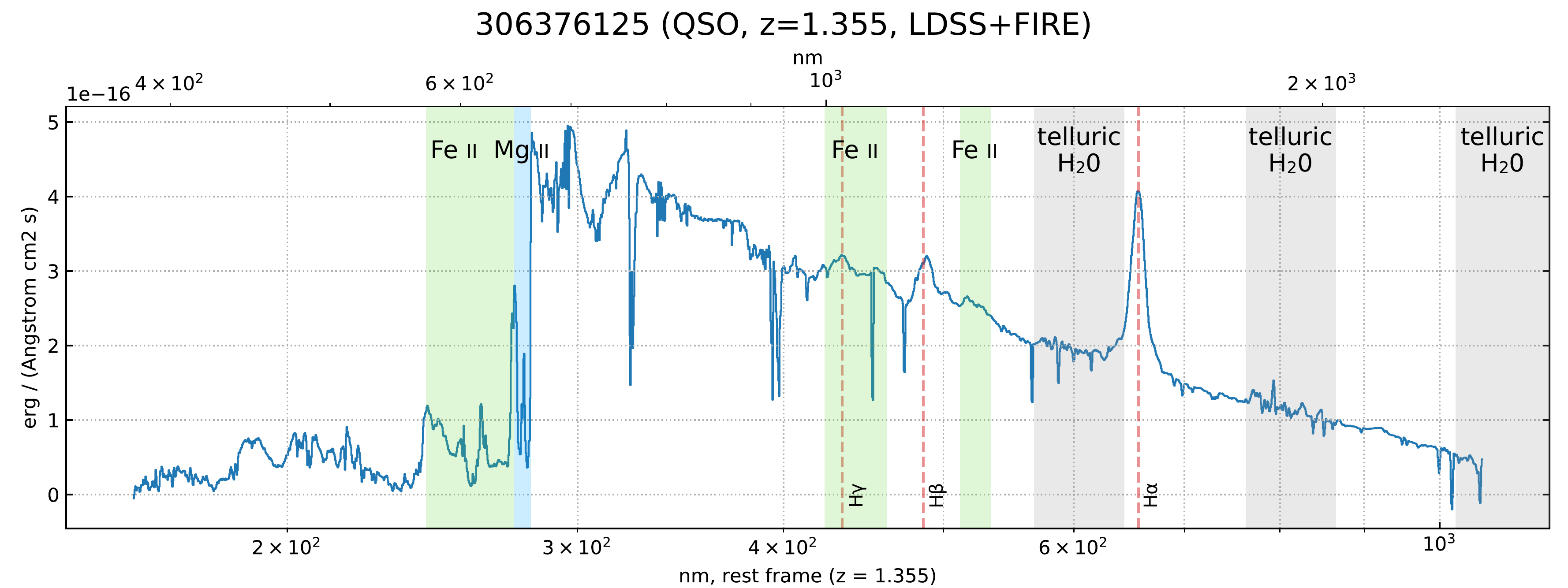}\\
\caption{Spectrum, resampled at 250 km/s, of a candidate identified as an OFeLoBAL QSO, with some spectral features 
highlighted. AGN emission lines and metal bands are shown in color, while telluric bands (which have been corrected for 
in data reduction) are shown in grey.}
\label{Fig:felobal}
\end{figure*}

Both in the pilot and the main campaign, we have encountered sources with pronounced absorption troughs whose 
classifications, due to the lack of strong emission lines,
is not straightforward. A fraction of these sources can in fact be classified as Overlapping Iron Low-ionisation 
BAL QSOs or OFeLoBAL \citep{Hazard+87}. OFeLoBALs are characterized by extensive systems of low ionization absorption 
which sometimes appear as broad saturated troughs (e.g.~\citealt{Hall+02}). The troughs may overlap to nearly 
completely absorb the continuum emission shortward of 2800 \AA, effectively mimicking the appearance of the Lyman 
forest. The estimated fraction of OFeLoBAL QSOs is around 2\% \citep{dai+12} and their number in our catalogue can 
be as high as the number of genuine $z>4$ QSOs.

The nature of these sources can be confirmed by identifying the Balmer series in the AGN emission, and in 
particular the H$\alpha$ emission line at 6563 \AA\ rest-frame. For $z\gtrsim 0.5$ this line is observed in 
the NIR band. We thus obtained NIR spectroscopic data for some of these sources, using the Folded-port InfraRed 
Echellete (FIRE), at the Baade Magellan telescope, with a long-slit configuration. 

A composite LDSS-3/FIRE spectrum of one such source (Skymapper ID 306376125), flux-calibrated in the visual band 
using an additional low-resolution spectrum taken by the Magellan Echellette Spectrograph (MagE) at Baade, is shown in 
figure \ref{Fig:felobal}. For this particular object, the H$\alpha$ and H$\beta$ line are clearly 
identified at $z=1.355 \pm 0.001$, with the possible additional detection of H$\gamma$ blended within a 
Fe \textsc{ii} emission complex. An overlapping Fe \textsc{ii}-Mg \textsc{ii} absorption complex at the 
same redshift is also clearly observed around 2600-2800 \AA\  rest-frame, corroborating the identification of 
the candidate as a OFeLoBAL QSO. A detailed analysis of this object and of other similar sources (currently assigned a 
flagB classification and not included in Tab.\ref{tab:NewSpec}), is beyond the scope of this work, and will be discussed '
in a future publication. 

\begin{figure*}[]
  \plotone{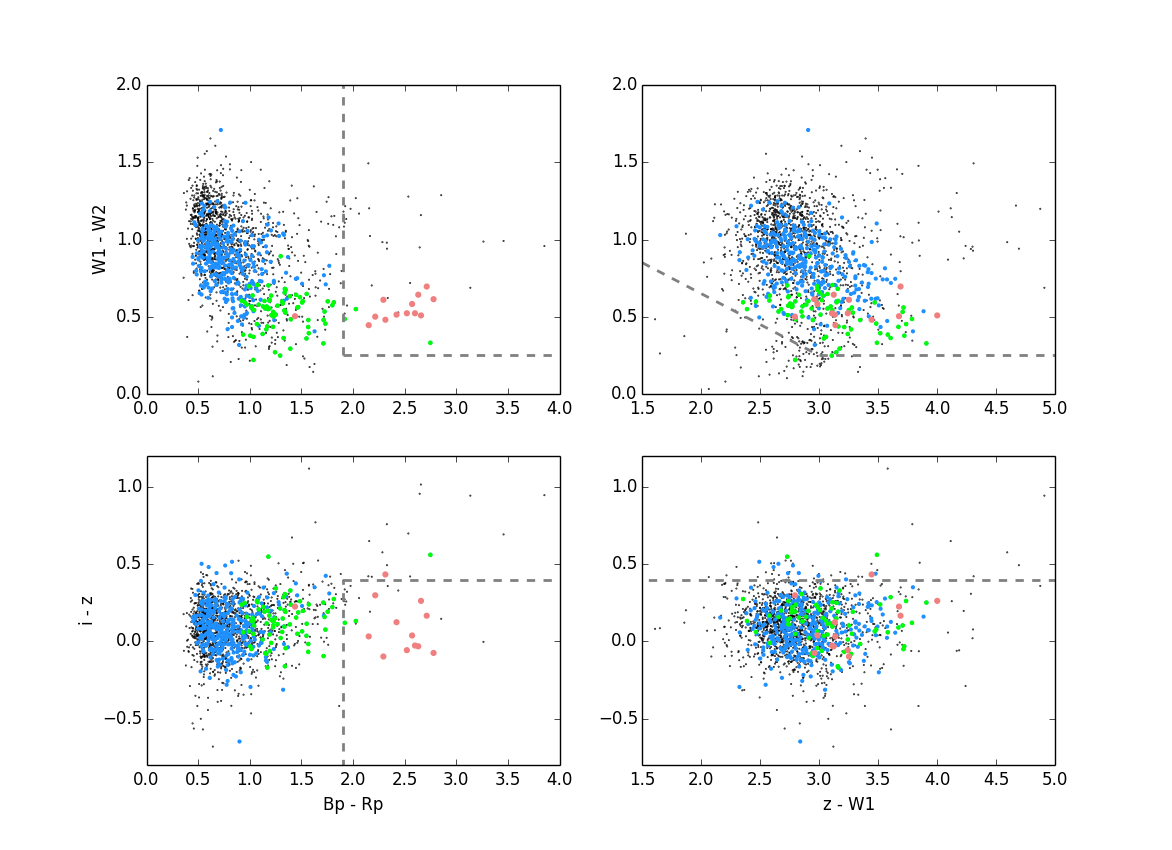}
  \caption{Color-color diagrams of all candidates with known redshift. Blue symbols are sources with $z \ge 2.5$, 
  green with $z \ge 3.5$ and red with $z \ge 4.5$. The lines correspond to cuts similar to the ones applied 
  by \cite{Wolf19} to select high redshift bright QSOs ($z \ge 4.5$). 
  The area delimited by the cuts is where the majority of $z \ge 4.5$ sources are found. WISE magnitudes are in 
  Vega and the rest are in the AB system.}
\label{Fig:cc}
\end{figure*}
\section{Properties of confirmed QSOs}
\subsection{IR colors }
Several recent surveys targeting super bright high-redshift QSOs are using WISE colors for the 
selection \citep{Wolf18,Sch19b,Wang16}. Based on a study by \citet{Wu12}, WISE colors are very good 
at distinguishing QSOs from late-type 
stars and more than half of SDSS QSOs have $W1-W2>0.57$. \cite{Wang16} showed that high redshift 
QSOs, $z>4.5$, are within the $0<(W1-W2)<1.5$ color range. 
This has also been confirmed by the survey conducted by \cite{Wolf18}. In Fig.\ref{Fig:cc} we show 
the distribution of our confirmed QSO 
candidates in the color space used by \cite{Wolf18}. In that work they were only 
interested in $z>4.5$ QSOs, but in our sample we have a wide range of redshifts going from $2.5\leq z<5.0 $. 
In order to be able to directly compare our distributions with the aforementioned works in the literature, 
presented WISE colors are in Vega magnitudes.

As can be seen in our plots, while all our candidates are indeed within the W1-W2[0,2] range, based only on 
IR colors it is difficult to distinguish between the various redshifts, since all sources are mixed in a small 
area. Instead, when we include the GAIA ($B_{p}-R_{p}$) color, the redshift groups are better separated as seen 
in the two panels on the left of the figure. Most of our $z \geq 4.5$ confirmed QSOs have a 
color of $B_{p}-R_{p}>$1.9, in line with the \cite{Wolf18} selection. In addition, all $z \geq 2.5$ QSOs 
have a color $B_{p}-R_{p}>$0.5, with the $z \geq 3.5$ source predominantly being in the area above $B_{p}-R_{p}=$1.0. 
Thus, including bluer colors in the selection, can help to better target higher redshift sources. 

\subsection{Crossmatch to Galex} \label{sec:galexCrossmatch}
We have cross-matched the QSOs found in QUBRICS with the catalogs of the sources detected by 
GALEX\footnote{All the GALEX data used in this section can be found 
in MAST: \dataset[https://doi:10.17909/T9H59D]{https://doi.org/10.17909/T9H59D}.} \citep{GALEX}. 
Of the 414 QSOs/AGN confirmed in Paper I and II (including those from table \ref{tab:AddSpectra}) 
we have 161 detections with a confidence level $> 2 \sigma$ in the NUV and 83 detections in the FUV; 75 objects 
are detected both in the FUV and NUV.
If we consider only QSOs with $z \geq 2.5$ ($z \geq 3.5$) out of 230 (54) objects 38 (3) have been detected in 
the NUV, 17 (2) in the FUV and 13 (1) both in the FUV and NUV.

In the designing phase of the QUBRICS we decided against using 
GALEX data for the selection of the candidates in order to avoid undesired selection effects relative to the 
fluctuations in the galactic absorption, which are stronger at UV wavelengths.
In fact, we plan to use the QUBRICS sample to study the HeII reionization and the cosmic UV background. In order 
to fulfill this goal, we need an unbiased sampling of all the possible lines-of-sight in our QSO catalogue and a UV 
selection may select preferentially less absorbed ones (e.g. \cite{Prochaska09}), which might bias future results 
towards shorter mean free paths in the IGM \citep{Romano19}.

Indeed, one of the QSOs confirmed in Paper I, J045011.37-432429.7, with $z=3.95$, shows a detection in the FUV and 
a non-detection in the NUV. 
FUV-loud QSOs are extremely rare at $z>3.5$ and precious to study the He II reionization \citep{Wor19}. 
J045011.37-432429.7 is potentially a case of transmission peaks that might indicate an example of patchy 
reionization of He II at $z \gtrsim 3.5$.

\begin{figure}[]
\plotone{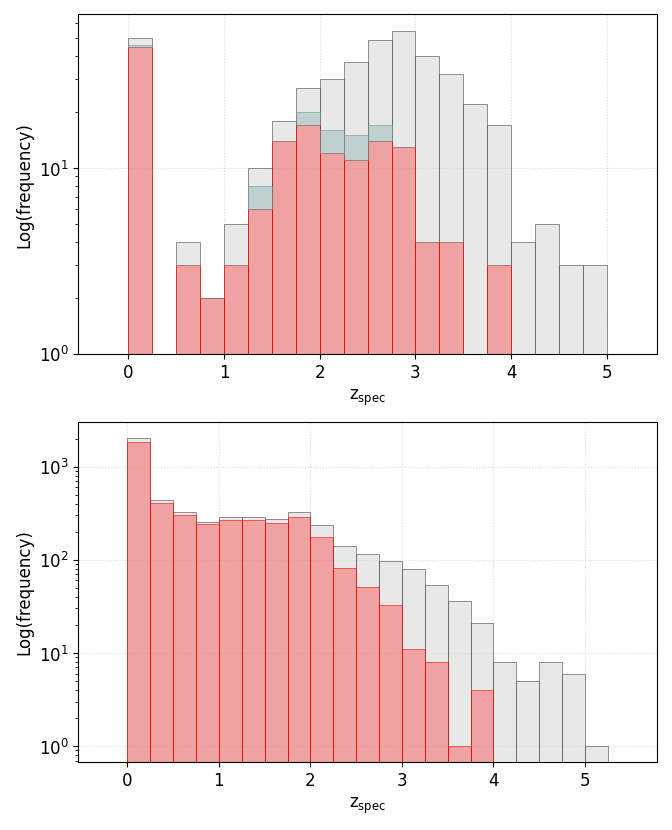}\\
\caption{Upper panel: redshift distribution of the 414 confirmed QSOs/AGN (grey columns): 169 of those have 
a corresponding GALEX source in FUV, NUV or both. 152 of them (red columns) were observed by us, while the 
remaining 17 sources (light blue columns) were found in the literature (\S \ref{ref:other_surv}). Lower panel: 
redshift distribution of QSOs/AGN in the QUBRICS {\it main sample}. Grey columns show the distribution for 
all the objects, red columns for those with a GALEX counterpart.}
\label{Fig:galexZ}
\end{figure}

As shown in Fig. \ref{Fig:galexZ}, all of QSOs/AGN with a GALEX counterpart have $z<4$, J045011.37-432429.7 
being the object with the highest recorded redshift. Moreover, considering the confirmed QSOs/AGN in 
paper I and II, 82\% of objects below $z=2$ have a GALEX counterpart. This number decreases drastically when 
considering higher redshift intervals: for $2 < z < 3$ ($3 < z < 4$) 35\% (11\%) of QSOs have a 
corresponding GALEX source. Similar results are obtained using all the QSOs/AGN known in the 
QUBRICS {\it main sample}, the biggest difference being for $2 < z < 3$ where a higher percentage (58\%) of 
sources have a GALEX counterpart.
In principle, the GALEX detection can be used as a veto criterium
to increase the efficiency in the selection of bright QSOs at $z>4$.

\begin{deluxetable*}{lcccccccr}
\tablecaption{The golden sample of Southern QSOs for the Sandage Test \label{tab:gold}}
\tablewidth{700pt}
\tabletypesize{\scriptsize}
\tablehead{
\colhead{Name} & 
\colhead{R.A. (J2000)} & 
\colhead{DEC. (J2000)} & 
\colhead{$m_r$} & 
\colhead{$m_i$} & 
\colhead{$G_{\rm GAIA}$} & 
\colhead{redshift} & 
\colhead{t(22.5 cm/s)} & 
\colhead{ID} \\
\colhead{} & 
\colhead{(deg)} & 
\colhead{(deg)} & 
\colhead{(mag)} & 
\colhead{(mag)} & 
\colhead{(mag)} & 
\colhead{} & 
\colhead{(h)} & 
\colhead{Skymapper}
} 
\startdata
J000322.94-260317.9 & 000.845639 & -26.055083 & 17.472 & 17.071 & 17.6019 & 4.111 & 61.2 &   7437380 \\ 
J000736.55-570151.8 & 001.902387 & -57.031067 & 17.992 & 17.240 & 17.9440 & 4.260 & 70.5 & 317358346 \\ 
J004131.41-493611.7 & 010.381000 & -49.603247 & 16.736 & 16.580 & 16.7048 & 3.240 & 68.5 & 317850840 \\ 
J010318.05-130509.9 & 015.825313 & -13.086163 & 17.637 & 17.242 & 17.7349 & 4.072 & 71.9 &   8430815 \\ 
J015558.26-192848.8 & 028.992816 & -19.480298 & 17.673 & 17.393 & 17.7969 & 3.655 & 96.7 &   8789744 \\ 
J015644.67-692216.1 & 029.186291 & -69.371155 & 16.637 & 16.325 & 16.5949 & 2.800 & 78.7 & 314938059 \\ 
J020413.26-325122.8 & 031.055213 & -32.856350 & 17.276 & 17.068 & 17.3900 & 3.835 & 66.2 &   6932623 \\ 
J030722.89-494548.2 & 046.845395 & -49.763410 &   *    & 17.407 & 18.4534 & 4.728 & 78.8 & 318204033 \\ 
J033015.31-543021.1 & 052.563812 & -54.505886 & 17.203 & 17.174 & 17.2208 & 3.400 & 97.4 & 316834279 \\ 
J042214.78-384452.9 & 065.561596 & -38.748007 & 16.896 & 16.765 & 16.9504 & 3.123 & 84.2 &  10914737 \\ 
J054803.19-484813.1 & 087.013348 & -48.803654 & 17.215 & 16.886 & 17.2414 & 4.147 & 51.4 & 316292063 \\ 
J093542.69-065118.8 & 143.927899 & -06.855260 & 18.008 & 17.424 & 18.1236 & 4.040 & 85.3 &  56483517 \\ 
J094253.51-110425.9 & 145.722949 & -11.073870 & 16.915 & 16.733 & 16.8174 & 3.093 & 87.0 &  56438607 \\ 
J101529.36-121314.3 & 153.872359 & -12.220648 & 17.857 & 17.255 & 17.7070 & 4.190 & 72.0 &  57929040 \\ 
J104856.82-163709.2 & 162.236717 & -16.619244 & 17.194 & 17.164 & 17.2249 & 3.370 & 98.0 &  57339247 \\ 
J105122.70-065047.8 & 162.844537 & -06.846624 & 17.584 & 17.345 & 17.8003 & 3.765 & 88.1 &  58181076 \\ 
J105449.68-171107.3 & 163.706991 & -17.185386 & 17.104 & 17.107 & 17.2681 & 3.765 & 70.8 &  57368436 \\ 
J111054.67-301129.8 & 167.727860 & -30.191652 &   *    & 17.346 & 18.5117 & 4.780 & 74.2 &  54680559 \\ 
J132029.98-052335.1 & 200.124872 & -05.393160 & 17.559 & 17.444 & 17.8589 & 3.700 & 99.4 &  65911949 \\ 
J144943.17-122717.5 & 222.429861 & -12.454868 & 16.953 & 16.703 & 17.0963 & 3.273 & 82.3 &  66962683 \\ 
J150527.83-204534.9 & 226.365934 & -20.759752 & 16.966 & 16.955 & 16.9888 & 3.090 & 91.3 &  72002629 \\ 
J162116.92-004250.8 & 245.320512 & -00.714142 & 17.457 & 17.386 & 17.6713 & 3.703 & 94.1 & 114286192 \\ 
J195302.67-381548.3 & 298.261202 & -38.263482 & 17.724 & 17.305 & 17.8896 & 3.712 & 87.0 & 135100950 \\ 
J200324.11-325145.0 & 300.850485 & -32.862537 & 17.470 & 17.296 & 17.6287 & 3.783 & 83.6 & 135386798 \\ 
J201741.49-281630.0 & 304.422893 & -28.274988 & 17.822 & 17.388 & 17.7996 & 3.685 & 95.0 & 136198132 \\ 
J212518.38-420547.6 & 321.326583 & -42.096609 & 17.266 & 17.363 & 17.6598 & 3.549 & 98.8 & 307443251 \\ 
J212540.96-171951.3 & 321.420690 & -17.330947 & 16.644 & 16.548 & 16.8730 & 3.897 & 39.9 &   2379862 \\ 
J212912.18-153840.9 & 322.300733 & -15.644734 & 17.035 & 16.824 & 17.0243 & 3.280 & 88.4 &   2552049 \\ 
J215228.21-444603.9 & 328.117538 & -44.767760 & 17.310 & 17.291 & 17.3645 & 3.473 & 94.8 & 307536968 \\ 
J215728.21-360215.1 & 329.367623 & -36.037557 &   *    & 17.367 & 18.2864 & 4.771 & 75.7 &    397340 \\ 
\enddata
\end{deluxetable*}

\section{A Golden Sample for the Sandage Test}
\label{sec:SandageT}

The redshifts of all cosmologically distant sources are expected to experience a small, systematic drift as 
a function of time due to the evolution of the expansion rate of the Universe \citep{Sandage62}. 
\cite{Liske08}, using extensive Monte Carlo simulations, determined the accuracy with which the redshift drift 
can be measured from the Lyman forest (and metal lines), $\sv$, as a function of signal-to-noise ratio and redshift:

\begin{equation}
\sv = g \times 1.35 
\left(\frac{{S/N}}{3350}\right)^{-1} 
\left(\frac{1 + \zqso}{5}\right)^{-\gamma} 
\left(\frac{\nqso}{30}\right)^{-0.5}
{\rm cm/s}
\label{sveqmne}
\end{equation}
where the symbol `S/N' refers to the total S/N per $0.0125$~\AA\ pixel per
object accumulated over all observations, $\nqso$ is the number of QSOs in the sample, $\zqso$ is the 
redshift of the QSO, the $\gamma$ exponent is $1.7$ for $\zqso \leq 4$ and 0.9 above. The form 
factor $g$ is equal to $1$ if all the targets are
observed twice, at the beginning and at the end of the experiment, and becomes larger if the 
measurements are distributed in time, reaching $1.7$ for a uniform distribution.
The S/N per pixel for photon-noise limited observations can be written as:

\begin{equation}
\label{sn}
{S/N} = 650 \left[ \frac{Z_X}{Z_r} \; 10^{0.4 (16 -
    m_X)} \; \left( \frac{D}{39 {\rm m}} \right)^2 \; \frac{\ti}{10 h} \; 
    \frac{\epsilon}{0.25} \right]^\frac{1}{2}
\end{equation}
where $D$, $\ti$ and $\epsilon$ are the telescope diameter, total
integration time and total efficiency, $Z_X$ and $m_X$ are the zeropoint
and apparent magnitude of the source in the $X$-band, respectively,
and $Z_r = (8.88 \times 10^{10})$~s$^{-1}$~m$^{-2}$~$\mu$m$^{-1}$ is
the AB zeropoint for an effective wavelength of
$6170$~\AA\ [corresponding to the Sloan Digital Sky Survey (SDSS)
$r$-band]. The normalisation of the above equation assumes a pixel
size of $0.0125$~\AA\ and a central
obscuration of the telescope's primary collecting area of $10$ per
cent.

\cite{Liske08} concluded that an ELT-type telescope (at the time planned with a 42m primary mirror) 
would be capable of unambiguously detecting the redshift drift over a period of $\sim 20$ yr using $4000$ h of 
observing time.
The estimated amount of time would obviously increase assuming the 39m primary mirror of the ELT and 
considering only the QSOs observable with the ELT,
typically those in the Southern hemisphere.
We have repeated the estimate of the time requested to carry out the Sandage Test with the 39m ELT, 
adopting a strategy that maximizes the significance of the detection of a non zero redshift 
drift \citep{Liske08} (i.e. with QSOs in the redshift range $2.8 \leq  z < 5$), aiming at a $3 \sigma$ 
detection and observing 30 targets twice at $25$ years distance.
Other assumptions are the same as in \cite{Liske08} except for the spectral slope of the QSO continuum, 
updated to be $f_{\lambda} \propto \lambda^{-1.3}$ \citep{Cristiani16}. For simplicity we have required 
that the spectra of all the
objects are integrated for a sufficiently long exposure time (different for each target, depending on 
its magnitude and redshift) to reach the same
velocity accuracy ($22.8$ cm/s) required for a global $3 \sigma$ detection of the drift.

We collect in Tab.~\ref{tab:gold} our proposed sample of 30 Southern QSOs, 
most suitable for the Sandage Test. 
Thanks to the detection of new bright QSOs
at high redshift, the total time required to carry out the Sandage Test turns out to be less than 2500 h 
and each QSO in Tab.~\ref{tab:gold} needs less than 100h of integration to provide a velocity accuracy of $22.8$ cm/s. 

\section{Discussion and Conclusions}

In this paper, we present spectroscopic identifications of QSO candidates in the QUBRICS sample, using our 
most updated selection criteria based on the CCA approach, which has been described in Paper I.
At present, after 26 observing runs at intermediate and large telescopes (duPont, Magellan, NTT, TNG), we have been 
able to approximately double the number of bright QSOs ($i\le 18.0$)
at $z \ge 2.5$ known in the Southern Hemisphere, bringing it to 428.
In this way it has been possible to relieve the persistent lack of bright targets for high resolution spectrographs 
of Southern observatories.

Using the new spectroscopic sample of bright QSOs from QUBRICS, we are able to further refine our selection criteria, 
by means of a new CCA training set. The completeness of the selection criterion, evaluated against the presently known 
bright QSOs at $z \geq 2.5 $, turns out to be higher than 90\%, while
the success rate is around 70\%. 
The success of the QUBRICS survey is particularly evident looking at Fig. \ref{Fig:i_vs_z}, where among the new identifications 
we show the two brightest QSOs at $z\ge 3.8$ in the
Southern hemisphere, with an increase of a factor of $\sim 3$ of the number of QSOs at $z\ge 3$ and $i\le 17.5$.

The QUBRICS survey also selects rare and peculiar sources, with strong absorption features, e.g. the OFeLoBAL QSO discussed in
section 4.3. With the progress of our survey it will be important to quantify the number densities of these peculiar 
sources and to compare them with the relative low incidence in the past surveys.

Due to the characteristics of our selection, QUBRICS is expected to be incomplete for QSOs at $z \gtrsim 4.5$ or with 
images distorted by lensing. In the future, we will address these issues and try to reduce the biases with improved, 
less stringent  selection criteria.

Surveys similar to QUBRICS have been started recently, based mainly on IR selections of bright high-z 
QSOs \citep[e.g.][]{Wolf19,Sch19b}.
IR colors from 2MASS and WISE are fundamental to distinguish high-z QSOs
from local galaxies or stars, but, taken alone, they are not optimal
for a photometric redshift refinement. As discussed in Section 5, the optical colors $Bp-Rp$ or $i-z$ are very 
useful to separate QSOs at $z\sim 2$ from the ones at $z\sim 4$. 

Besides, adopting further filters, in particular at short wavelengths (e.g. GALEX), it would be possible to further 
improve the success rate of the survey at high z. However, due to the possibility of introducing a bias against clear 
lines of sight in the intervening IGM, which can mimic the photometry of a lower redshift source \citep{Romano19},
we chose not to use the GALEX photometry for the selection. 
In this way, among the newly discovered QSOs at $z>2.5$, we have 29 objects with FUV or NUV detections, potentially 
important for follow up studies with COS for the HeII reionization \citep[e.g.][]{Wor16,Wor19}.

The adopted CCA method to compute a photometric redshift, described in Paper I, is based on SEDs extending from 
the u band to the WISE bands, and should be less affected by possible biases due to the presence of significant 
Lyman limit absorbers with short mean free paths, which are evident at $z\sim 3.5$ on QSO surveys based on $u-g$ color 
selections, as shown by \citet{Prochaska09, Cristiani16, Romano19}. In the future, we will address the effects of 
optical color selections on the QSO spectral properties and Lyman limit statistics.

The discovery of bright cosmic beacons is especially important for the study of the IGM at high-z.
In particular, the QUBRICS survey provides a large sample of bright high-z QSOs for the Sandage Test with future 40 
meter class telescopes. We estimate that at present, less than 2500 hours of observations in 25 years are needed with 
the ELT-HIRES to carry out the redshift drift measurement \citep{Liske08} at the precision required to 
have a 3$\sigma$ detection of the cosmological signal. Before QUBRICS,
the targets available in the Southern Hemisphere would have required, in the same metrics, about 4000 hours to 
accomplish this goal. QUBRICS, in a sense, contributes to a multi-million saving, considering the projected cost of 
a night at ELT ($\gtrsim 50$ MEU or $\gtrsim 60$ M\$ at the assumed cost of 320 KEU for an ELT observing night and 
an average of 9 hours per night). 

We are continuing the pursuit for the brightest cosmic beacons, exploring other innovative methods, with the main 
goal of giving a realistic possibility to execute the Sandage test, only dreamed of a few years ago.

\acknowledgments
We thank Gabor Worseck for enlightening discussions.

This work is based on data products from observations made with ESO Telescopes at La Silla Paranal Observatory 
under ESO programmes ID 103.A-0746(A), 0103.A-0746(B), and 0104.A-0754(A).

The national facility capability for SkyMapper has been funded through ARC LIEF grant LE130100104 from the 
Australian Research Council, awarded to the University of Sydney, the Australian National University, 
Swinburne University of Technology, the University of Queensland, the University of Western Australia, 
the University of Melbourne, Curtin University of Technology, Monash University and the Australian 
Astronomical Observatory. SkyMapper is owned and operated by The Australian National University's Research School 
of Astronomy and Astrophysics. The survey data were processed and provided by the SkyMapper Team at ANU. 
The SkyMapper node of the All-Sky Virtual Observatory (ASVO) is hosted at the National Computational 
Infrastructure (NCI). Development and support the SkyMapper node of the ASVO has been funded in part by 
Astronomy Australia Limited (AAL) and the Australian Government through the Commonwealth's Education 
Investment Fund (EIF) and National Collaborative Research Infrastructure Strategy (NCRIS), particularly 
the National eResearch Collaboration Tools and Resources (NeCTAR) and the Australian National Data Service Projects (ANDS)

This work has made use of data from the European Space Agency (ESA) mission {\it Gaia} (\url{https://www.cosmos.esa.int/gaia}), 
processed by the {\it Gaia} Data Processing and Analysis 
Consortium (DPAC, \url{https://www.cosmos.esa.int/web/gaia/dpac/consortium}). 
Funding for the DPAC has been provided by national institutions, in particular the institutions participating 
in the {\it Gaia} Multilateral Agreement.

This publication makes use of data products from the Two Micron All Sky Survey, which is a joint project of 
the University of Massachusetts and the Infrared Processing and Analysis Center/California Institute of Technology, 
funded by the National Aeronautics and Space Administration and the National Science Foundation

This publication makes use of data products from the Wide-field Infrared Survey Explorer, which is a joint 
project of the University of California, Los Angeles, and the Jet Propulsion Laboratory/California Institute 
of Technology, funded by the National Aeronautics and Space Administration.

This paper includes data gathered with the 6.5 meter Magellan Telescopes located at Las Campanas Observatory, Chile.

This research is based on observations made with the Galaxy Evolution Explorer, obtained from the MAST data 
archive at the Space Telescope Science Institute, which is operated by the Association of Universities for 
Research in Astronomy, Inc., under NASA contract NAS 5–26555.

We thank Societ\`a Astronomica Italiana (SAIt), Ennio Poretti, Gloria Andreuzzi, Marco Pedani, Vittoria 
Altomonte and Andrea Cama for the observation support at TNG.  Part of the observations discussed in this 
work are based on observations made with the Italian Telescopio Nazionale Galileo (TNG) operated on the 
island of La Palma by the Fundacion Galileo Galilei of the INAF (Istituto Nazionale di Astrofisica) at the 
Spanish Observatorio del Roque de los Muchachos of the Instituto de Astrofisica de Canarias.

\facilities{Skymapper, Wise, 2MASS, Gaia, NTT (EFOSC2), Magellan:Baade (IMACS), Magellan:Clay (LDSS-3), 
Magellan:Baade (FIRE), du Pont (WFCCD), TNG (Dolores)}

\appendix

\section{CCA matrixes}
Our QSO candidate selection procedure relies on a CCA transformation aimed at predicting an object 
type classification for each source, based on the photometric magnitudes in the bands where these 
estimates are available. In particular we associate a discrete numeric label to all sources with 
known object type classification, and search for the transformation matrix to be applied to the 
available magnitudes which maximizes the correlation with the above mentioned label. Note that the 
above process identifies the best possible linear transformation while the best performances would 
likely be obtained with a non-linear one, although the latter can not be obtained analytically. In 
order to improve our classification algorithm we also included the logarithm of each magnitude estimate 
in the CCA analysis. It should be noted that the availability of magnitude estimates is not the same for 
all the sources in our main sample, hence we identified 34 sub-samples with homogeneous availability and 
repeated the above process for each of them, resulting in 34 different CCA matrices. The largest sub-sample 
covers 43\% of our main sample (438687 sources, with 58581 sources lacking a known object type classification). 
In Tab.~\ref{tab:CCA_matrix} we report the transformation matrix used to calculate the CCA 
coordinate (i.e., the value displayed on the vertical axis of Fig.~\ref{Fig:cca1} in both Paper I 
and Paper II), for the above mentioned sub-sample.

\begin{deluxetable}{lcc}
\tablecaption{CCA transformation matrix for the magnitude band combination in the largest 
sub-sample (covering 43\% of the main sample).}
\label{tab:CCA_matrix}
\tablewidth{700pt}
\tabletypesize{\scriptsize}
\tablehead{
\colhead{Band} & \colhead{Paper I} & \colhead{This work}\\
}
\startdata
G (Gaia)            &    -9.652   &     -10.02  \\
BP (Gaia)           &    -4.571   &     -3.680  \\
RP (Gaia)           &    +10.21   &     +9.332  \\
g (Skymapper)       &    +2.297   &     +2.091  \\
r (Skymapper)       &    +4.637   &     +4.006  \\
i (Skymapper)       &    -2.129   &     -1.386  \\
z (Skymapper)       &    +3.089   &     +3.407  \\
J (2MASS)           &   -0.8026   &    -0.7531  \\
H (2MASS)           &    +3.733   &     +3.984  \\
K (2MASS)           &    +0.458   &    +0.5796  \\
W1 (WISE)           &    -8.250   &     -7.629  \\
W2 (WISE)           &    +0.263   &    -0.7174  \\
W3 (WISE)           &    +2.348   &     +2.326  \\
log G (Gaia)        &    +142.9   &     +149.4  \\
log BP (Gaia)       &    +72.69   &     +57.33  \\
log RP (Gaia)       &    -138.6   &     -125.4  \\
log g (Skymapper)   &    -40.05   &     -36.85  \\
log r (Skymapper)   &    -71.36   &     -61.19  \\
log i (Skymapper)   &    +33.78   &     +22.85  \\
log z (Skymapper)   &    -45.00   &     -49.95  \\
log J (2MASS)       &    +14.76   &     +13.89  \\
log H (2MASS)       &    -50.95   &     -53.82  \\
log K (2MASS)       &    -8.637   &     -10.47  \\
log W1 (WISE)       &    +141.7   &     +132.2  \\
log W2 (WISE)       &    -22.17   &     -6.822  \\
log W3 (WISE)       &    -41.23   &     -40.81  \\
offset              &    +10.45   &     +5.014  \\
\enddata
\end{deluxetable}

All magnitudes should be converted to the AB system before applying the transformation, and an 
offset (last line of Tab.~\ref{tab:CCA_matrix}) should be added to the resulting label to obtain 
a classification on the same scale of the training. We can provide the matrices for the remaining 
sources on request.\footnote{Please write to giorgio.calderone@inaf.it.} Further details are available in Paper I.

Note that the presence of strong correlations among the magnitudes in different bands prevents us 
from drawing any conclusion by just comparing the numbers in the above matrices. The overall performance 
of each matrix, in terms of success rates and completeness, should thus be estimated by comparing the 
results with the test dataset, as discussed in \S\ref{sec:selection}.


\bibliography{ReferencesPapII}{}

\begin{thebibliography}{}
\expandafter\ifx\csname natexlab\endcsname\relax\def\natexlab#1{#1}\fi
\providecommand{\url}[1]{\href{#1}{#1}}
\providecommand{\dodoi}[1]{doi:~\href{http://doi.org/#1}{\nolinkurl{#1}}}
\providecommand{\doeprint}[1]{\href{http://ascl.net/#1}{\nolinkurl{http://ascl.net/#1}}}
\providecommand{\doarXiv}[1]{\href{https://arxiv.org/abs/#1}{\nolinkurl{https://arxiv.org/abs/#1}}}

\bibitem[{{Anderson}(1984)}]{CCA}
{Anderson}, T. 1984, {An introduction to multivariate statistical analysis}
  (Wiley Ed.)

\bibitem[{{Bate} {et~al.}(2007){Bate}, {Webster}, \& {Wyithe}}]{Bate07}
{Bate}, N.~F., {Webster}, R.~L., \& {Wyithe}, J.~S.~B. 2007, \mnras, 381, 1591,
  \dodoi{10.1111/j.1365-2966.2007.12330.x}

\bibitem[{{Becker} {et~al.}(2019){Becker}, {D'Aloisio}, {Davies}, {Hennawi}, \&
  {Simcoe}}]{Beck19}
{Becker}, G., {D'Aloisio}, A., {Davies}, F.~B., {Hennawi}, J.~F., \& {Simcoe},
  R.~A. 2019, \baas, 51, 440.
\newblock \doarXiv{1903.05199}

\bibitem[{{Bianchi} {et~al.}(2017){Bianchi}, {Shiao}, \& {Thilker}}]{GALEX}
{Bianchi}, L., {Shiao}, B., \& {Thilker}, D. 2017, \apjs, 230, 24,
  \dodoi{10.3847/1538-4365/aa7053}

\bibitem[{{Bonvin} {et~al.}(2017){Bonvin}, {Courbin}, {Suyu}, {Marshall},
  {Rusu}, {Sluse}, {Tewes}, {Wong}, {Collett}, {Fassnacht}, {Treu}, {Auger},
  {Hilbert}, {Koopmans}, {Meylan}, {Rumbaugh}, {Sonnenfeld}, \&
  {Spiniello}}]{Bonvin17}
{Bonvin}, V., {Courbin}, F., {Suyu}, S.~H., {et~al.} 2017, \mnras, 465, 4914,
  \dodoi{10.1093/mnras/stw3006}

\bibitem[{{Calderone} {et~al.}(2019){Calderone}, {Boutsia}, {Cristiani},
  {Grazian}, {Amorin}, {D{\textquoteright}Odorico}, {Cupani}, {Fontanot}, \&
  {Salvato}}]{calderone19}
{Calderone}, G., {Boutsia}, K., {Cristiani}, S., {et~al.} 2019, \apj, 887, 268,
  \dodoi{10.3847/1538-4357/ab510a}

\bibitem[{{Colless} {et~al.}(2001){Colless}, {Dalton}, {Maddox}, {Sutherland },
  {Norberg}, {Cole}, {Bland -Hawthorn}, {Bridges}, {Cannon}, {Collins},
  {Couch}, {Cross}, {Deeley}, {De Propris}, {Driver}, {Efstathiou}, {Ellis},
  {Frenk}, {Glazebrook}, {Jackson}, {Lahav}, {Lewis}, {Lumsden}, {Madgwick},
  {Peacock}, {Peterson}, {Price}, {Seaborne}, \& {Taylor}}]{Colless01}
{Colless}, M., {Dalton}, G., {Maddox}, S., {et~al.} 2001, \mnras, 328, 1039,
  \dodoi{10.1046/j.1365-8711.2001.04902.x}

\bibitem[{{Cooke} \& {Fumagalli}(2018)}]{Cooke18}
{Cooke}, R.~J., \& {Fumagalli}, M. 2018, Nature Astronomy, 2, 957,
  \dodoi{10.1038/s41550-018-0584-z}

\bibitem[{{Cristiani} {et~al.}(2016){Cristiani}, {Serrano}, {Fontanot},
  {Vanzella}, \& {Monaco}}]{Cristiani16}
{Cristiani}, S., {Serrano}, L.~M., {Fontanot}, F., {Vanzella}, E., \& {Monaco},
  P. 2016, \mnras, 462, 2478, \dodoi{10.1093/mnras/stw1810}

\bibitem[{{Dai} {et~al.}(2012){Dai}, {Shankar}, \& {Sivakoff}}]{dai+12}
{Dai}, X., {Shankar}, F., \& {Sivakoff}, G.~R. 2012, \apj, 757, 180,
  \dodoi{10.1088/0004-637X/757/2/180}

\bibitem[{{D'Odorico} {et~al.}(2016){D'Odorico}, {Cristiani}, {Pomante},
  {Carswell}, {Viel}, {Barai}, {Becker}, {Calura}, {Cupani}, {Fontanot},
  {Haehnelt}, {Kim}, {Miralda-Escud{\'e}}, {Rorai}, {Tescari}, \&
  {Vanzella}}]{Dod16}
{D'Odorico}, V., {Cristiani}, S., {Pomante}, E., {et~al.} 2016, \mnras, 463,
  2690, \dodoi{10.1093/mnras/stw2161}

\bibitem[{{Fan} \& {SDSS Collaboration}(2000)}]{fan2000}
{Fan}, X., \& {SDSS Collaboration}. 2000, in American Astronomical Society
  Meeting Abstracts, Vol. 197, American Astronomical Society Meeting Abstracts,
  27.01

\bibitem[{{Fontanot} {et~al.}(2007){Fontanot}, {Cristiani}, {Monaco}, {Nonino},
  {Vanzella}, {Brandt}, {Grazian}, \& {Mao}}]{Fontanot07}
{Fontanot}, F., {Cristiani}, S., {Monaco}, P., {et~al.} 2007, \aap, 461, 39,
  \dodoi{10.1051/0004-6361:20066073}

\bibitem[{{Fontanot} {et~al.}(2020){Fontanot}, {De Lucia}, {Hirschmann}, {Xie},
  {Monaco}, {Menci}, {Fiore}, {Feruglio}, {Cristiani}, \& {Shankar}}]{Fon2020}
{Fontanot}, F., {De Lucia}, G., {Hirschmann}, M., {et~al.} 2020, arXiv
  e-prints, arXiv:2002.10576.
\newblock \doarXiv{2002.10576}

\bibitem[{{Gaia Collaboration} {et~al.}(2018){Gaia Collaboration}, {Brown},
  {Vallenari}, {Prusti}, {de Bruijne}, {Babusiaux}, {Bailer-Jones}, {Biermann},
  {Evans}, {Eyer}, \& et~al.}]{Gaia18}
{Gaia Collaboration}, {Brown}, A.~G.~A., {Vallenari}, A., {et~al.} 2018, \aap,
  616, A1, \dodoi{10.1051/0004-6361/201833051}

\bibitem[{{Grazian} {et~al.}(2018){Grazian}, {Giallongo}, {Boutsia},
  {Cristiani}, {Vanzella}, {Scarlata}, {Santini}, {Pentericci}, {Merlin},
  {Menci}, {Fontanot}, {Fontana}, {Fiore}, {Civano}, {Castellano}, {Brusa},
  {Bonchi}, {Carini}, {Cusano}, {Faccini}, {Garilli}, {Marchetti}, {Rossi}, \&
  {Speziali}}]{Grazia18}
{Grazian}, A., {Giallongo}, E., {Boutsia}, K., {et~al.} 2018, \aap, 613, A44,
  \dodoi{10.1051/0004-6361/201732385}

\bibitem[{{Hall} {et~al.}(2002){Hall}, {Anderson}, {Strauss}, {York},
  {Richards}, {Fan}, {Knapp}, {Schneider}, {Vanden Berk}, {Geballe}, {Bauer},
  {Becker}, {Davis}, {Rix}, {Nichol}, {Bahcall}, {Brinkmann}, {Brunner},
  {Connolly}, {Csabai}, {Doi}, {Fukugita}, {Gunn}, {Haiman}, {Harvanek},
  {Heckman}, {Hennessy}, {Inada}, {Ivezi{\'c}}, {Johnston}, {Kleinman},
  {Krolik}, {Krzesinski}, {Kunszt}, {Lamb}, {Long}, {Lupton}, {Miknaitis},
  {Munn}, {Narayanan}, {Neilsen}, {Newman}, {Nitta}, {Okamura}, {Pentericci},
  {Pier}, {Schlegel}, {Snedden}, {Szalay}, {Thakar}, {Tsvetanov}, {White}, \&
  {Zheng}}]{Hall+02}
{Hall}, P.~B., {Anderson}, S.~F., {Strauss}, M.~A., {et~al.} 2002, \apjs, 141,
  267, \dodoi{10.1086/340546}

\bibitem[{{Hazard} {et~al.}(1987){Hazard}, {McMahon}, {Webb}, \&
  {Morton}}]{Hazard+87}
{Hazard}, C., {McMahon}, R.~G., {Webb}, J.~K., \& {Morton}, D.~C. 1987, \apj,
  323, 263, \dodoi{10.1086/165823}

\bibitem[{{Ir{\v{s}}i{\v{c}}} {et~al.}(2017){Ir{\v{s}}i{\v{c}}}, {Viel},
  {Haehnelt}, {Bolton}, {Cristiani}, {Becker}, {D'Odorico}, {Cupani}, {Kim},
  {Berg}, {L{\'o}pez}, {Ellison}, {Christensen}, {Denney}, \&
  {Worseck}}]{Irsic17}
{Ir{\v{s}}i{\v{c}}}, V., {Viel}, M., {Haehnelt}, M.~G., {et~al.} 2017, \prd,
  96, 023522, \dodoi{10.1103/PhysRevD.96.023522}

\bibitem[{{Jones} {et~al.}(2009){Jones}, {Read}, {Saunders}, {Colless},
  {Jarrett}, {Parker}, {Fairall}, {Mauch}, {Sadler}, {Watson}, {Burton},
  {Campbell}, {Cass}, {Croom}, {Dawe}, {Fiegert}, {Frankcombe}, {Hartley},
  {Huchra}, {James}, {Kirby}, {Lahav}, {Lucey}, {Mamon}, {Moore}, {Peterson},
  {Prior}, {Proust}, {Russell}, {Safouris}, {Wakamatsu}, {Westra}, \&
  {Williams}}]{6dF09}
{Jones}, D.~H., {Read}, M.~A., {Saunders}, W., {et~al.} 2009, \mnras, 399, 683,
  \dodoi{10.1111/j.1365-2966.2009.15338.x}

\bibitem[{{Leite} {et~al.}(2016){Leite}, {Martins}, {Molaro}, {Corre}, \&
  {Cristiani}}]{Leite16}
{Leite}, A.~C.~O., {Martins}, C.~J.~A.~P., {Molaro}, P., {Corre}, D., \&
  {Cristiani}, S. 2016, \prd, 94, 123512, \dodoi{10.1103/PhysRevD.94.123512}

\bibitem[{{Liske} {et~al.}(2008){Liske}, {Grazian}, {Vanzella}, {Dessauges},
  {Viel}, {Pasquini}, {Haehnelt}, {Cristiani}, {Pepe}, {Avila}, {Bonifacio},
  {Bouchy}, {Dekker}, {Delabre}, {D'Odorico}, {D'Odorico}, {Levshakov},
  {Lovis}, {Mayor}, {Molaro}, {Moscardini}, {Murphy}, {Queloz}, {Shaver},
  {Udry}, {Wiklind}, \& {Zucker}}]{Liske08}
{Liske}, J., {Grazian}, A., {Vanzella}, E., {et~al.} 2008, \mnras, 386, 1192,
  \dodoi{10.1111/j.1365-2966.2008.13090.x}

\bibitem[{{P{\^a}ris} {et~al.}(2018){P{\^a}ris}, {Petitjean}, {Aubourg},
  {Myers}, {Streblyanska}, {Lyke}, {Anderson}, {Armengaud}, {Bautista},
  {Blanton}, {Blomqvist}, {Brinkmann}, {Brownstein}, {Brand t}, {Burtin},
  {Dawson}, {de la Torre}, {Georgakakis}, {Gil-Mar{\'\i}n}, {Green}, {Hall},
  {Kneib}, {LaMassa}, {Le Goff}, {MacLeod}, {Mariappan}, {McGreer}, {Merloni},
  {Noterdaeme}, {Palanque-Delabrouille}, {Percival}, {Ross}, {Rossi},
  {Schneider}, {Seo}, {Tojeiro}, {Weaver}, {Weijmans}, {Y{\`e}che}, {Zarrouk},
  \& {Zhao}}]{Paris18}
{P{\^a}ris}, I., {Petitjean}, P., {Aubourg}, {\'E}., {et~al.} 2018, \aap, 613,
  A51, \dodoi{10.1051/0004-6361/201732445}

\bibitem[{{Prochaska} {et~al.}(2009){Prochaska}, {Worseck}, \&
  {O'Meara}}]{Prochaska09}
{Prochaska}, J.~X., {Worseck}, G., \& {O'Meara}, J.~M. 2009, \apjl, 705, L113,
  \dodoi{10.1088/0004-637X/705/2/L113}

\bibitem[{{Romano} {et~al.}(2019){Romano}, {Grazian}, {Giallongo}, {Cristiani},
  {Fontanot}, {Boutsia}, {Fiore}, \& {Menci}}]{Romano19}
{Romano}, M., {Grazian}, A., {Giallongo}, E., {et~al.} 2019, \aap, 632, A45,
  \dodoi{10.1051/0004-6361/201935550}

\bibitem[{{Sandage}(1962)}]{Sandage62}
{Sandage}, A. 1962, \apj, 136, 319, \dodoi{10.1086/147385}

\bibitem[{{Schindler} {et~al.}(2019{\natexlab{a}}){Schindler}, {Fan},
  {McGreer}, {Yang}, {Wang}, {Green}, {Fynbo}, {Krogager}, {Green}, {Huang},
  {Kadowaki}, {Patej}, {Wu}, \& {Yue}}]{Sch19a}
{Schindler}, J.-T., {Fan}, X., {McGreer}, I.~D., {et~al.} 2019{\natexlab{a}},
  \apj, 871, 258, \dodoi{10.3847/1538-4357/aaf86c}

\bibitem[{{Schindler} {et~al.}(2019{\natexlab{b}}){Schindler}, {Fan}, {Huang},
  {Yue}, {Yang}, {Hall}, {Wenzl}, {Hughes}, {Litke}, \& {Rees}}]{Sch19b}
{Schindler}, J.-T., {Fan}, X., {Huang}, Y.-H., {et~al.} 2019{\natexlab{b}},
  \apjs, 243, 5, \dodoi{10.3847/1538-4365/ab20d0}

\bibitem[{{Skrutskie} {et~al.}(2006){Skrutskie}, {Cutri}, {Stiening},
  {Weinberg}, {Schneider}, {Carpenter}, {Beichman}, {Capps}, {Chester},
  {Elias}, {Huchra}, {Liebert}, {Lonsdale}, {Monet}, {Price}, {Seitzer},
  {Jarrett}, {Kirkpatrick}, {Gizis}, {Howard}, {Evans}, {Fowler}, {Fullmer},
  {Hurt}, {Light}, {Kopan}, {Marsh}, {McCallon}, {Tam}, {Van Dyk}, \&
  {Wheelock}}]{Ref2MASS}
{Skrutskie}, M.~F., {Cutri}, R.~M., {Stiening}, R., {et~al.} 2006, \aj, 131,
  1163, \dodoi{10.1086/498708}

\bibitem[{{Tie} {et~al.}(2017){Tie}, {Martini}, {Mudd}, {Ostrovski}, {Reed},
  {Lidman}, {Kochanek}, {Davis}, {Sharp}, {Uddin}, {King}, {Wester}, {Tucker},
  {Tucker}, {Buckley-Geer}, {Carollo}, {Childress}, {Glazebrook}, {Hinton},
  {Lewis}, {Macaulay}, {O'Neill}, {Abbott}, {Abdalla}, {Annis},
  {Benoit-L{\'e}vy}, {Bertin}, {Brooks}, {Carnero Rosell}, {Carrasco Kind},
  {Carretero}, {Cunha}, {da Costa}, {DePoy}, {Desai}, {Doel}, {Eifler},
  {Evrard}, {Finley}, {Flaugher}, {Fosalba}, {Frieman}, {Garc{\'\i}a-Bellido},
  {Gaztanaga}, {Gerdes}, {Goldstein}, {Gruen}, {Gruendl}, {Gutierrez},
  {Honscheid}, {James}, {Kuehn}, {Kuropatkin}, {Lima}, {Maia}, {Marshall},
  {Menanteau}, {Miller}, {Miquel}, {Nichol}, {Nord}, {Ogando}, {Plazas},
  {Romer}, {Sanchez}, {Santiago}, {Scarpine}, {Schubnell}, {Sevilla-Noarbe},
  {Smith}, {Soares-Santos}, {Sobreira}, {Suchyta}, {Swanson}, {Tarle},
  {Thomas}, {Walker}, \& {DES Collaboration}}]{DES_17}
{Tie}, S.~S., {Martini}, P., {Mudd}, D., {et~al.} 2017, \aj, 153, 107,
  \dodoi{10.3847/1538-3881/aa5b8d}

\bibitem[{{Valiante} {et~al.}(2016){Valiante}, {Schneider}, {Volonteri}, \&
  {Omukai}}]{Vali16}
{Valiante}, R., {Schneider}, R., {Volonteri}, M., \& {Omukai}, K. 2016, \mnras,
  457, 3356, \dodoi{10.1093/mnras/stw225}

\bibitem[{{V{\'e}ron-Cetty} \& {V{\'e}ron}(2010)}]{Veron10}
{V{\'e}ron-Cetty}, M.~P., \& {V{\'e}ron}, P. 2010, \aap, 518, A10,
  \dodoi{10.1051/0004-6361/201014188}

\bibitem[{{Wang} {et~al.}(2016){Wang}, {Wu}, {Fan}, {Yang}, {Yi}, {Bian},
  {McGreer}, {Yang}, {Ai}, {Dong}, {Zuo}, {Jiang}, {Green}, {Wang}, {Cai},
  {Wang}, \& {Yue}}]{Wang16}
{Wang}, F., {Wu}, X.-B., {Fan}, X., {et~al.} 2016, \apj, 819, 24,
  \dodoi{10.3847/0004-637X/819/1/24}

\bibitem[{{Webster} {et~al.}(1991){Webster}, {Ferguson}, {Corrigan}, \&
  {Irwin}}]{Webster91}
{Webster}, R.~L., {Ferguson}, A.~M.~N., {Corrigan}, R.~T., \& {Irwin}, M.~J.
  1991, \aj, 102, 1939, \dodoi{10.1086/116015}

\bibitem[{{Werk} {et~al.}(2014){Werk}, {Prochaska}, {Tumlinson}, {Peeples},
  {Tripp}, {Fox}, {Lehner}, {Thom}, {O'Meara}, {Ford}, {Bordoloi}, {Katz},
  {Tejos}, {Oppenheimer}, {Dav{\'e}}, \& {Weinberg}}]{Werk14}
{Werk}, J.~K., {Prochaska}, J.~X., {Tumlinson}, J., {et~al.} 2014, \apj, 792,
  8, \dodoi{10.1088/0004-637X/792/1/8}

\bibitem[{{Wolf} {et~al.}(2018){Wolf}, {Onken}, {Luvaul}, {Schmidt}, {Bessell},
  {Chang}, {Da Costa}, {Mackey}, {Martin-Jones}, {Murphy}, {Preston}, {Scalzo},
  {Shao}, {Smillie}, {Tisserand}, {White}, \& {Yuan}}]{Wolf18}
{Wolf}, C., {Onken}, C.~A., {Luvaul}, L.~C., {et~al.} 2018, \pasa, 35, e010,
  \dodoi{10.1017/pasa.2018.5}

\bibitem[{{Wolf} {et~al.}(2020){Wolf}, {Hon}, {Bian}, {Onken}, {Alonzi},
  {Bessell}, {Li}, {Schmidt}, \& {Tisserand}}]{Wolf19}
{Wolf}, C., {Hon}, W.~J., {Bian}, F., {et~al.} 2020, \mnras, 491, 1970,
  \dodoi{10.1093/mnras/stz2955}

\bibitem[{{Worseck} {et~al.}(2019){Worseck}, {Davies}, {Hennawi}, \&
  {Prochaska}}]{Wor19}
{Worseck}, G., {Davies}, F.~B., {Hennawi}, J.~F., \& {Prochaska}, J.~X. 2019,
  \apj, 875, 111, \dodoi{10.3847/1538-4357/ab0fa1}

\bibitem[{{Worseck} {et~al.}(2016){Worseck}, {Prochaska}, {Hennawi}, \&
  {McQuinn}}]{Wor16}
{Worseck}, G., {Prochaska}, J.~X., {Hennawi}, J.~F., \& {McQuinn}, M. 2016,
  \apj, 825, 144, \dodoi{10.3847/0004-637X/825/2/144}

\bibitem[{{Worseck} {et~al.}(2014){Worseck}, {Prochaska}, {O'Meara}, {Becker},
  {Ellison}, {Lopez}, {Meiksin}, {M{\'e}nard}, {Murphy}, \&
  {Fumagalli}}]{Worseck14}
{Worseck}, G., {Prochaska}, J.~X., {O'Meara}, J.~M., {et~al.} 2014, \mnras,
  445, 1745, \dodoi{10.1093/mnras/stu1827}

\bibitem[{{Wright} {et~al.}(2010){Wright}, {Eisenhardt}, {Mainzer}, {Ressler},
  {Cutri}, {Jarrett}, {Kirkpatrick}, {Padgett}, {McMillan}, {Skrutskie},
  {Stanford}, {Cohen}, {Walker}, {Mather}, {Leisawitz}, {Gautier}, {McLean},
  {Benford}, {Lonsdale}, {Blain}, {Mendez}, {Irace}, {Duval}, {Liu}, {Royer},
  {Heinrichsen}, {Howard}, {Shannon}, {Kendall}, {Walsh}, {Larsen}, {Cardon},
  {Schick}, {Schwalm}, {Abid}, {Fabinsky}, {Naes}, \& {Tsai}}]{Wri10}
{Wright}, E.~L., {Eisenhardt}, P. R.~M., {Mainzer}, A.~K., {et~al.} 2010, \aj,
  140, 1868, \dodoi{10.1088/0004-6256/140/6/1868}

\bibitem[{{Wu} {et~al.}(2012){Wu}, {Hao}, {Jia}, {Zhang}, \& {Peng}}]{Wu12}
{Wu}, X.-B., {Hao}, G., {Jia}, Z., {Zhang}, Y., \& {Peng}, N. 2012, \aj, 144,
  49, \dodoi{10.1088/0004-6256/144/2/49}

\bibitem[{{Wu} {et~al.}(2015){Wu}, {Wang}, {Fan}, {Yi}, {Zuo}, {Bian}, {Jiang},
  {McGreer}, {Wang}, {Yang}, {Yang}, {Thompson}, \& {Beletsky}}]{Wu15}
{Wu}, X.-B., {Wang}, F., {Fan}, X., {et~al.} 2015, \nat, 518, 512,
  \dodoi{10.1038/nature14241}

\end{thebibliography}

\begin{longrotatetable}
\begin{deluxetable*}{lccccccccc}
\tablecaption{All observed sources with secure redshift identification \label{chartable}}
\label{tab:NewSpec}
\tablewidth{700pt}
\tabletypesize{\scriptsize}
\tablehead{
\colhead{Skymapper} & \colhead{R.A. (J2000)} & \colhead{DEC. (J2000)} & 
\colhead{$m_i$} & \colhead{$z_{\rm spec}$} &  \colhead{Class} & 
\colhead{Candidate} & \colhead{Candidate} & 
\colhead{Date-Obs} & \colhead{Inst} \\ 
\colhead{ID} & \colhead{} & \colhead{} &
\colhead{(mag)} & \colhead{} & \colhead{} & 
\colhead{PaperI} & \colhead{New} &
\colhead{} & \colhead{} \\
} 
\startdata
107306    &  21:01:34.72 &  -30:10:29.75 &  17.616  &    2.761 &    QSO    & Y & Y &  2019-09 &  NTT \\
343536    &  21:38:41.84 &  -33:49:37.21 &  17.660  &    3.253 &    QSO    & Y & Y &  2019-09 &  NTT \\
397340$^{a}$    &  21:57:28.21 &  -36:02:15.11 &  17.367  &    4.771 &    QSO    & Y & Y &  2019-08 &  WFCCD \\
583628$^{b}$    &  21:32:25.90 &  -28:31:33.24 &  17.605  &    2.821 &    QSO    & Y & Y &  2019-09 &  NTT \\
755996    &  21:30:54.92 &  -22:46:54.51 &  17.693  &    1.749 &    QSO    & Y & N &  2019-09 &  NTT \\
838086    &  21:49:28.24 &  -27:47:46.22 &  17.114  &    2.495 &    QSO    & Y & Y &  2019-09 &  NTT \\
1316873   &  22:55:26.10 &  -36:39:33.80 &  17.531  &    1.148 &    QSO    & Y & N &  2019-09 &  NTT \\
1318863   &  22:52:47.11 &  -35:42:01.46 &  17.692  &    2.630 &    QSO    & Y & Y &  2019-09 &  NTT \\
1826062   &  23:01:39.67 &  -28:59:46.08 &  17.062  &    0.000 &    STAR   & Y & N &  2019-08 &  WFCCD \\
2415223   &  21:05:01.83 &  -13:34:40.76 &  17.514  &    2.346 &    QSO    & Y & Y &  2019-09 &  NTT \\
2809723   &  21:54:54.84 &  -18:06:03.36 &  17.690  &    3.000 &    QSO    & Y & Y &  2019-09 &  NTT \\
3310797   &  21:32:17.64 &  -06:07:50.64 &  17.732  &    0.000 &    STAR   & Y & N &  2019-09 &  NTT \\
3479216   &  21:19:19.03 &  -04:56:56.93 &  16.959  &    2.672 &    QSO    & Y & Y &  2019-09 &  NTT \\
3709817   &  21:40:53.51 &  -05:56:47.09 &  17.582  &    2.001 &    QSO    & Y & Y &  2019-09 &  NTT \\
4045023   &  21:55:13.29 &  -03:16:05.61 &  17.410  &    3.690 &    QSO    & Y & Y &  2019-08 &  WFCCD \\
4342067   &  22:57:00.78 &  -21:42:09.66 &  17.280  &    2.833 &    QSO    & Y & Y &  2019-09 &  WFCCD \\
4342686$^{b}$   &  22:59:39.04 &  -22:50:34.90 &  17.709  &    3.486 &    QSO    & Y & Y &  2019-09 &  NTT \\
4441259   &  22:27:54.04 &  -15:49:38.61 &  17.379  &    2.647 &    QSO    & Y & Y &  2019-09 &  NTT \\
4538557   &  22:46:46.21 &  -16:02:41.71 &  17.547  &    2.259 &    QSO    & Y & Y &  2019-09 &  NTT \\
4788903   &  23:56:40.97 &  -21:15:56.84 &  16.764  &    1.738 &    QSO    & Y & Y &  2019-09 &  NTT \\
4922910   &  23:43:15.99 &  -18:21:12.21 &  17.617  &    3.332 &    QSO    & Y & Y &  2019-09 &  NTT \\
4924968   &  23:40:28.39 &  -17:24:46.12 &  17.609  &    4.257 &    QSO    & Y & Y &  2019-09 &  NTT \\
5003358   &  23:59:21.09 &  -14:57:22.91 &  17.705  &    2.953 &    QSO    & Y & Y &  2019-09 &  NTT \\
5035730   &  22:22:41.07 &  -10:27:09.45 &  17.613  &    1.820 &    QSO    & Y & Y &  2019-09 &  NTT \\
5081975   &  22:23:13.66 &  -08:54:16.34 &  17.593  &    2.706 &    QSO    & Y & Y &  2019-09 &  NTT \\
5552633   &  23:15:12.19 &  -11:23:43.85 &  17.704  &    2.561 &    QSO    & Y & Y &  2019-09 &  NTT \\
5564557   &  23:23:46.94 &  -12:15:03.37 &  17.599  &    2.752 &    QSO    & Y & Y &  2019-09 &  NTT \\
5931917   &  23:47:49.34 &  -04:33:32.60 &  17.679  &    1.880 &    QSO    & Y & N &  2019-12 &  WFCCD \\
6009251   &  00:01:43.86 &  -44:25:35.10 &  17.624  &    2.572 &    QSO    & Y & Y &  2019-09 &  NTT \\
6146221   &  00:21:33.83 &  -37:40:08.04 &  17.641  &    2.573 &    QSO    & Y & Y &  2019-09 &  NTT \\
6232292   &  01:11:58.59 &  -40:33:17.64 &  17.611  &    3.432 &    QSO    & Y & Y &  2019-09 &  NTT \\
6419289   &  00:00:11.42 &  -33:10:50.86 &  17.605  &    3.106 &    QSO    & Y & Y &  2019-09 &  NTT \\
6427176   &  00:15:48.69 &  -34:03:50.83 &  17.507  &    0.055 &    AGN    & Y & Y &  2019-09 &  NTT \\
6810884   &  01:43:22.60 &  -25:38:34.27 &  17.748  &    2.307 &    QSO    & Y & Y &  2019-09 &  NTT \\
6932623   &  02:04:13.26 &  -32:51:22.80 &  17.068  &    3.835 &    QSO    & Y & Y &  2019-11 &  WFCCD \\
7015154   &  02:39:54.21 &  -34:38:36.83 &  17.911  &    0.000 &    STAR   & Y & N &  2019-09 &  NTT \\
7088896   &  02:49:54.68 &  -33:48:35.91 &  16.706  &    2.712 &    QSO    & Y & Y &  2019-09 &  NTT \\
7125196   &  02:44:06.47 &  -30:07:49.93 &  17.703  &    0.063 &    AGN    & Y & Y &  2019-09 &  NTT \\
7134124   &  02:57:16.05 &  -31:26:44.44 &  17.671  &    3.599 &    QSO    & Y & Y &  2019-09 &  NTT \\
7152915   &  01:51:52.22 &  -32:52:38.96 &  17.206  &    0.196 &    AGN    & Y & N &  2019-08 &  WFCCD \\
7250804$^{b}$   &  01:50:41.58 &  -25:08:46.32 &  17.603  &    3.596 &    QSO    & Y & Y &  2019-09 &  NTT \\
7277955   &  02:20:23.13 &  -26:12:40.04 &  17.405  &    3.096 &    QSO    & Y & Y &  2019-11 &  LDSS-3 \\
7292553   &  02:09:46.00 &  -24:01:11.19 &  17.679  &    1.370 &    QSO    & Y & N &  2019-09 &  NTT \\
7341051   &  02:44:49.14 &  -29:04:49.25 &  17.422  &    3.240 &    QSO    & Y & Y &  2019-11 &  WFCCD \\
7560979   &  00:32:03.76 &  -21:27:29.73 &  17.685  &    1.855 &    QSO    & Y & N &  2019-12 &  WFCCD \\
7565232$^{c}$   &  00:39:01.09 &  -21:44:28.89 &  17.597  &    3.027 &    QSO    & Y & Y &  2019-09 &  NTT \\
7687616   &  00:39:19.11 &  -19:06:19.36 &  17.745  &    0.852 &    QSO    & Y & Y &  2019-09 &  NTT \\
7751454   &  00:44:18.35 &  -15:04:13.35 &  17.442  &    2.755 &    QSO    & Y & Y &  2019-08 &  WFCCD \\
7829753   &  01:27:48.47 &  -23:34:58.12 &  17.176  &    2.933 &    QSO    & Y & Y &  2019-08 &  WFCCD \\
7901686   &  01:10:02.61 &  -18:43:20.07 &  17.561  &    2.940 &    QSO    & Y & Y &  2019-09 &  NTT \\
7931324   &  01:02:53.62 &  -14:15:22.53 &  17.482  &    0.035 &    GAL    & Y & N &  2019-08 &  WFCCD \\
7957603   &  01:33:16.52 &  -17:39:23.61 &  17.663  &    2.903 &    QSO    & Y & Y &  2019-09 &  NTT \\
8145069$^{b}$   &  00:43:46.84 &  -11:17:01.77 &  17.510  &    3.510 &    QSO    & Y & Y &  2019-09 &  NTT \\
8170168   &  00:32:20.84 &  -08:13:00.92 &  17.356  &    2.838 &    QSO    & Y & Y &  2019-11 &  WFCCD \\
8430815$^{b}$   &  01:03:18.06 &  -13:05:09.89 &  17.242  &    4.072 &    QSO    & Y & Y &  2019-08 &  WFCCD \\
8493702$^{c}$   &  01:20:35.95 &  -09:46:32.02 &  17.653  &    3.009 &    QSO    & Y & Y &  2019-09 &  NTT \\
8513332   &  01:34:10.33 &  -12:16:31.76 &  17.072  &    2.010 &    QSO    & Y & N &  2019-11 &  WFCCD \\
8529305   &  01:40:41.94 &  -12:36:22.39 &  17.587  &    3.178 &    QSO    & Y & Y &  2019-09 &  NTT \\
8590857   &  00:58:33.99 &  -04:08:23.66 &  17.393  &    0.048 &    AGN    & Y & Y &  2019-08 &  WFCCD \\
8619006$^{**}$   &  01:00:49.23 &  -03:19:13.97 &  17.25   &    3.45  &    QSO    & Y & ? &  2018-10  &   IMACS        \\
8733320   &  01:37:35.48 &  -02:46:06.09 &  17.678  &    0.048 &    AGN    & Y & Y &  2019-09 &  NTT \\
8789744$^{c}$   &  01:55:58.27 &  -19:28:48.98 &  17.393  &    3.655 &    QSO    & Y & Y &  2019-08 &  WFCCD \\
8795094   &  02:05:56.09 &  -19:53:25.57 &  17.149  &    0.205 &    AGN    & Y & N &  2019-11 &  LDSS-3 \\
8904188   &  02:16:41.24 &  -16:00:18.55 &  17.748  &    3.039 &    QSO    & Y & Y &  2019-09 &  NTT \\
8918253   &  02:20:15.17 &  -14:53:54.10 &  17.664  &    2.365 &    QSO    & Y & Y &  2019-12 &  LDSS-3 \\
9089958$^{**}$   &  02:45:43.76 &  -12:06:03.71 &  16.672  &    0.03  &    GAL    & N & ? &  2018-09  &    LDSS-3      \\
9177858   &  02:08:56.49 &  -09:01:06.21 &  17.733  &    2.665 &    QSO    & Y & Y &  2019-09 &  NTT \\
9250343$^{c}$   &  01:59:00.68 &  -03:27:37.18 &  17.479  &    2.946 &    QSO    & Y & Y &  2019-08 &  WFCCD \\
9407358   &  02:52:26.99 &  -09:29:18.21 &  17.063  &    2.255 &    QSO    & Y & Y &  2019-11 &  WFCCD \\
9892519   &  22:44:25.86 &  +02:10:08.30 &  17.015  &    0.060 &    GAL    & Y & N &  2019-09 &  NTT \\
10476365  &  03:36:43.14 &  -32:20:04.22 &  17.583  &    0.050 &    AGN    & Y & Y &  2019-09 &  NTT \\
10531556  &  04:08:12.64 &  -34:37:48.67 &  17.659  &    3.279 &    QSO    & Y & Y &  2020-01 &  NTT \\
10572165  &  03:16:42.30 &  -28:03:45.07 &  17.594  &    2.313 &    QSO    & Y & Y &  2019-09 &  NTT \\
10719380  &  03:56:17.49 &  -32:08:19.87 &  17.585  &    2.291 &    QSO    & Y & Y &  2019-09 &  NTT \\
10748545  &  03:55:08.17 &  -30:11:53.97 &  17.316  &    0.068 &    AGN    & Y & N &  2019-11 &  WFCCD \\
10790773  &  03:49:16.29 &  -26:55:39.63 &  17.674  &    1.938 &    QSO    & Y & N &  2019-09 &  NTT \\
10798423  &  03:36:20.00 &  -24:25:47.32 &  17.308  &    0.540 &    QSO    & N & N &  2019-12 &  WFCCD \\
10830553  &  03:53:01.84 &  -25:48:23.05 &  17.749  &    3.083 &    QSO    & Y & Y &  2019-09 &  NTT \\
10875592  &  04:05:48.52 &  -24:21:15.18 &  17.449  &    2.785 &    QSO    & Y & Y &  2019-12 &  WFCCD \\
11028492  &  04:21:11.68 &  -34:34:03.64 &  17.537  &    1.780 &    QSO    & Y & N &  2019-11 &  WFCCD \\
11053925$^{*}$  &  04:40:01.13 &  -37:55:34.52 &  18.317  &    3.852 &    QSO    & - & - &  2020-01 &  NTT \\
11163743  &  05:12:15.00 &  -41:20:13.86 &  17.616  &    3.144 &    QSO    & Y & Y &  2020-01 &  NTT \\
11270169$^{*}$  &  05:04:56.56 &  -39:52:45.03 &  18.251  &    3.237 &    QSO    & - & - &  2020-01 &  NTT \\
11343829  &  05:31:08.81 &  -37:48:20.40 &  17.651  &    3.170 &    QSO    & Y & Y &  2019-11 &  LDSS-3 \\
11382764  &  05:51:33.62 &  -40:56:31.83 &  17.879  &    2.819 &    QSO    & Y & Y &  2020-01 &  NTT \\
11477581  &  04:23:33.67 &  -32:54:14.19 &  17.463  &    2.270 &    QSO    & Y & Y &  2019-12 &  WFCCD \\
11513515  &  04:30:16.02 &  -31:57:32.97 &  17.918  &    3.552 &    QSO    & Y & Y &  2020-01 &  NTT \\
11530749$^{*}$  &  04:42:05.78 &  -34:46:49.78 &  18.358  &    2.959 &    QSO    & - & - &  2020-01 &  NTT \\
11621558$^{*}$  &  04:14:00.70 &  -27:20:58.62 &  18.268  &    3.998 &    QSO    & - & - &  2020-01 &  NTT \\
11625837$^{*}$  &  04:23:28.88 &  -27:52:23.74 &  18.166  &    3.016 &    QSO    & - & - &  2020-01 &  NTT \\
11639037  &  04:29:37.15 &  -28:35:43.10 &  17.389  &    2.569 &    QSO    & Y & Y &  2019-12 &  WFCCD \\
11708647  &  04:44:10.65 &  -29:13:07.15 &  17.476  &    1.841 &    QSO    & Y & Y &  2020-01 &  NTT \\
11797890  &  05:13:04.21 &  -35:28:51.50 &  17.397  &    3.050 &    QSO    & Y & Y &  2019-12 &  WFCCD \\
11871566$^{*}$  &  05:19:04.89 &  -33:03:27.23 &  18.031  &    3.272 &    QSO    & - & - &  2020-01 &  NTT \\
12020560  &  05:50:37.14 &  -32:39:11.73 &  17.913  &    3.369 &    QSO    & Y & Y &  2020-01 &  NTT \\
12185422  &  05:42:50.82 &  -31:03:24.12 &  17.634  &    2.620 &    QSO    & Y & Y &  2019-11 &  WFCCD \\
12375861  &  03:06:13.04 &  -15:17:46.67 &  17.329  &    2.590 &    QSO    & Y & Y &  2019-12 &  WFCCD \\
12411621  &  03:19:01.77 &  -17:07:19.05 &  17.626  &    2.526 &    QSO    & Y & Y &  2019-09 &  NTT \\
12432405  &  03:07:00.94 &  -14:04:56.95 &  17.474  &    1.670 &    QSO    & Y & Y &  2019-12 &  WFCCD \\
12455576$^{**}$  &  03:03:54.28 &  -10:50:49.01 &  16.35   &    0.04  &    GAL    & N & ? &  2018-09  &    LDSS-3      \\
12476780$^{*}$  &  03:19:04.82 &  -13:42:00.18 &  18.370  &    2.989 &    QSO    & - & - &  2020-01 &  NTT \\
12513469  &  03:29:03.97 &  -12:05:42.70 &  17.641  &    2.614 &    QSO    & Y & Y &  2019-09 &  NTT \\
12559809  &  03:42:21.23 &  -17:55:11.47 &  17.236  &    2.033 &    QSO    & Y & Y &  2019-11 &  LDSS-3 \\
12668545  &  03:41:38.05 &  -11:42:59.46 &  17.205  &    2.800 &    QSO    & Y & Y &  2019-12 &  WFCCD \\
12750382  &  04:09:04.91 &  -13:37:00.61 &  17.519  &    2.771 &    QSO    & Y & Y &  2020-01 &  NTT \\
12817400  &  03:26:45.28 &  -08:47:00.54 &  17.365  &    2.348 &    QSO    & Y & Y &  2019-12 &  WFCCD \\
12931253  &  03:23:14.75 &  -02:29:58.63 &  17.881  &    0.082 &    GAL    & Y & N &  2019-09 &  NTT \\
12941812  &  03:25:25.48 &  -02:18:04.20 &  17.523  &    2.268 &    QSO    & Y & N &  2019-12 &  LDSS-3 \\
12957365$^{*}$  &  03:31:06.08 &  -00:32:38.53 &  16.189  &    0.000 &    STAR   & - & - &  2020-01 &  LDSS-3 \\
12989144  &  03:48:51.22 &  -10:15:53.61 &  16.974  &    2.498 &    QSO    & Y & Y &  2019-12 &  WFCCD \\
13173703$^{*}$  &  03:36:34.04 &  -00:49:16.40 &  14.678  &    0.002 &    GAL    & - & - &  2020-01 &  NTT \\
13291946  &  04:09:34.40 &  -02:17:38.04 &  17.640  &    3.202 &    QSO    & Y & Y &  2020-01 &  NTT \\
13323971  &  04:17:23.77 &  -21:47:13.67 &  17.344  &    2.060 &    QSO    & Y & Y &  2019-12 &  WFCCD \\
13409036  &  04:37:47.11 &  -24:16:23.62 &  17.239  &    1.980 &    QSO    & N & N &  2019-12 &  WFCCD \\
13422105  &  04:38:56.76 &  -22:56:53.36 &  15.949  &    0.026 &    GAL    & N & N &  2019-12 &  WFCCD \\
13479091$^{*}$  &  04:39:54.55 &  -19:43:51.32 &  18.035  &    2.609 &    QSO    & - & - &  2020-01 &  NTT \\
13623074  &  04:40:53.44 &  -18:43:28.06 &  17.674  &    2.832 &    QSO    & Y & Y &  2019-11 &  LDSS-3 \\
13754735  &  05:04:50.03 &  -24:32:51.42 &  17.488  &    1.789 &    QSO    & Y & Y &  2019-12 &  LDSS-3 \\
14047570  &  05:37:28.45 &  -21:56:10.42 &  17.125  &    2.835 &    QSO    & Y & Y &  2019-12 &  WFCCD \\
14148404  &  05:14:11.75 &  -19:01:39.46 &  17.681  &    2.609 &    QSO    & Y & Y &  2019-11 &  WFCCD \\
14203137  &  05:24:38.81 &  -18:16:04.14 &  16.831  &    1.969 &    QSO    & Y & Y &  2019-12 &  WFCCD \\
14271415  &  05:25:49.05 &  -16:16:12.37 &  17.843  &    3.567 &    QSO    & Y & Y &  2020-01 &  NTT \\
14661311  &  04:50:19.69 &  -11:18:36.99 &  17.394  &    2.560 &    QSO    & Y & Y &  2019-12 &  WFCCD \\
14756860$^{*}$  &  05:03:29.07 &  -09:41:52.22 &  18.389  &    3.307 &    QSO    & - & - &  2020-01 &  NTT \\
14797495$^{c}$  &  04:25:03.70 &  -05:45:08.94 &  17.498  &    3.040 &    QSO    & Y & Y &  2019-11 &  WFCCD \\
14832947  &  04:32:49.02 &  -05:41:48.21 &  17.600  &    0.165 &    AGN    & Y & N &  2019-11 &  LDSS-3 \\
14921486  &  04:32:50.36 &  -01:12:53.72 &  16.690  &    1.691 &    QSO    & Y & N &  2019-09 &  NTT \\
14930439$^{c}$  &  04:36:23.92 &  -00:04:02.89 &  17.404  &    3.852 &    QSO    & Y & Y &  2020-01 &  NTT \\
15015396  &  04:58:13.47 &  -03:52:51.60 &  17.375  &    3.000 &    QSO    & Y & Y &  2019-12 &  WFCCD \\
15040824  &  04:46:17.76 &  -02:20:15.64 &  17.687  &    2.533 &    QSO    & Y & Y &  2019-11 &  LDSS-3 \\
15088950$^{*}$  &  05:02:54.73 &  -03:09:23.73 &  18.089  &    3.689 &    QSO    & - & - &  2020-01 &  NTT \\
15248403  &  05:07:32.14 &  -06:58:02.27 &  17.427  &    2.960 &    QSO    & Y & Y &  2019-12 &  WFCCD \\
16164525$^{*}$  &  06:01:37.14 &  -43:41:43.80 &  18.378  &    3.499 &    QSO    & - & - &  2020-01 &  NTT \\
16539422  &  06:02:18.73 &  -36:59:24.72 &  17.518  &    2.920 &    QSO    & Y & N &  2019-11 &  WFCCD \\
47669241$^{*}$  &  08:15:45.64 &  -01:38:18.59 &  16.261  &    0.000 &    STAR   & - & - &  2020-01 &  LDSS-3 \\
48196484$^{*}$  &  08:58:49.38 &  -02:34:05.48 &  16.381  &    0.000 &    STAR   & - & - &  2020-01 &  LDSS-3 \\
48449084$^{*}$  &  08:44:22.01 &  +04:32:44.27 &  18.036  &    3.327 &    QSO    & - & - &  2020-01 &  NTT \\
51718111$^{*}$  &  10:07:03.55 &  -23:57:13.36 &  18.287  &    3.347 &    QSO    & - & - &  2020-01 &  NTT \\
54192246  &  10:59:56.17 &  -30:28:27.95 &  17.487  &    2.610 &    QSO    & Y & Y &  2020-01/02 &  NTT/LDSS-3 \\
54298405  &  10:58:09.30 &  -26:05:42.29 &  17.505  &    2.876 &    QSO    & Y & Y &  2020-01 &  NTT \\
54680559$^{a}$  &  11:10:54.68 &  -30:11:29.88 &  17.346  &    4.779 &    QSO    & Y & Y &  2020-01 &  NTT \\
55462080  &  09:30:53.18 &  -12:12:29.67 &  17.007  &    0.000 &    STAR   & Y & N &  2020-02 &  LDSS-3 \\
55582991  &  09:49:02.92 &  -20:07:23.36 &  16.298  &    1.866 &    QSO    & N & N &  2020-02 &  LDSS-3 \\
55799640  &  09:44:35.17 &  -16:16:32.31 &  17.224  &    2.897 &    QSO    & Y & Y &  2020-02 &  LDSS-3 \\
56003412  &  10:12:57.45 &  -14:34:12.40 &  17.540  &    3.229 &    QSO    & Y & Y &  2020-01 &  NTT \\
56507631$^{*}$  &  09:49:34.23 &  -08:03:08.77 &  18.086  &    2.333 &    QSO    & - & - &  2020-01 &  NTT \\
56610095  &  09:59:20.61 &  -06:47:25.66 &  17.189  &    1.797 &    QSO    & Y & Y &  2020-02 &  LDSS-3 \\
56626491  &  10:04:16.57 &  -07:46:34.72 &  17.327  &    0.558 &    QSO    & Y & N &  2020-01 &  NTT \\
56648626  &  09:40:52.85 &  -05:09:32.44 &  17.144  &    2.370 &    QSO    & Y & Y &  2020-02 &  LDSS-3 \\
56660065  &  09:41:19.36 &  -04:08:12.54 &  16.754  &    0.000 &    STAR   & Y & N &  2020-01 &  LDSS-3 \\
56681692  &  09:43:51.97 &  -03:21:02.73 &  17.464  &    1.701 &    QSO    & N & N &  2020-02 &  LDSS-3 \\
56699168$^{*}$  &  09:37:47.61 &  -02:01:26.30 &  16.610  &    0.000 &    STAR   & - & - &  2020-01 &  LDSS-3 \\
56749347  &  09:52:29.66 &  -04:08:02.85 &  17.487  &    2.074 &    QSO    & Y & Y &  2020-01 &  LDSS-3 \\
56791679$^{*}$  &  09:54:32.83 &  -02:07:51.29 &  16.369  &    0.000 &    STAR   & - & - &  2020-01 &  LDSS-3 \\
56823181$^{*}$  &  10:12:05.23 &  -02:15:24.61 &  16.358  &    0.000 &    STAR   & - & - &  2020-01 &  LDSS-3 \\
56933068  &  10:25:07.94 &  -20:58:19.50 &  17.152  &    2.308 &    QSO    & Y & Y &  2020-02 &  LDSS-3 \\
57291610$^{*}$  &  10:32:52.81 &  -13:52:04.59 &  18.098  &    2.950 &    QSO    & - & - &  2020-01 &  NTT \\
57292353  &  10:35:25.96 &  -14:39:47.44 &  17.298  &    2.888 &    QSO    & Y & Y &  2020-01 &  NTT \\
57466047  &  11:13:57.01 &  -26:04:58.35 &  16.817  &    2.949 &    QSO    & Y & Y &  2020-01/02 &  NTT/LDSS-3 \\
57669548$^{*}$  &  11:34:40.56 &  -21:03:22.78 &  16.205  &    2.774 &    QSO    & - & - &  2020-01 &  NTT \\
58115878$^{*}$  &  10:54:45.47 &  -10:53:29.52 &  18.053  &    0.156 &    GAL    & - & - &  2020-01 &  LDSS-3 \\
58723356  &  11:13:32.47 &  -03:09:13.98 &  17.949  &    3.731 &    QSO    & Y & Y &  2020-01 &  NTT \\
58823120$^{*}$  &  11:57:43.62 &  -06:50:04.73 &  18.247  &    2.748 &    QSO    & - & - &  2020-01 &  NTT \\
60288193  &  13:04:42.64 &  -36:04:27.43 &  17.892  &    0.155 &    AGN    & Y & N &  2020-01 &  NTT \\
61068064  &  12:41:46.38 &  -29:34:49.05 &  17.404  &    3.048 &    QSO    & Y & Y &  2020-01 &  NTT \\
62209153  &  13:55:44.05 &  -35:08:15.11 &  17.447  &    0.000 &    STAR   & Y & N &  2020-01 &  NTT \\
63230995$^{b}$  &  14:08:01.83 &  -27:58:20.34 &  17.641  &    4.447 &    QSO    & Y & Y &  2020-01 &  NTT \\
63285615  &  13:49:40.58 &  -26:33:42.92 &  17.222  &    3.023 &    QSO    & Y & Y &  2020-01 &  NTT \\
63739700  &  14:32:28.52 &  -24:44:35.07 &  17.357  &    2.718 &    QSO    & Y & Y &  2020-01 &  NTT \\
63982611  &  12:15:37.66 &  -24:53:23.63 &  17.043  &    0.083 &    GAL    & N & N &  2020-01 &  NTT \\
64116716  &  12:49:08.53 &  -25:32:38.86 &  17.437  &    0.147 &    AGN    & N & N &  2020-01 &  LDSS-3 \\
65072412  &  12:27:24.77 &  -11:33:35.05 &  14.951  &    0.000 &    STAR   & N & N &  2020-01 &  NTT \\
65723003$^{*}$  &  13:20:01.62 &  -06:53:21.60 &  18.188  &    1.329 &    QSO    & - & - &  2020-01 &  NTT \\
65842041$^{*}$  &  13:45:17.91 &  -08:29:57.17 &  18.254  &    0.472 &    AGN    & - & - &  2020-01 &  NTT \\
98859792  &  15:43:22.37 &  -19:42:41.10 &  16.004  &    0.177 &    AGN    & Y & N &  2019-08 &  WFCCD \\
125383528 &  19:40:12.00 &  -41:18:22.33 &  17.642  &    2.618 &    QSO    & Y & Y &  2019-09 &  NTT \\
135055221 &  20:06:00.80 &  -39:37:32.61 &  17.623  &    1.340 &    QSO    & Y & N &  2019-09 &  NTT \\
135169820 &  20:19:01.86 &  -38:27:21.89 &  17.387  &    2.070 &    QSO    & Y & N &  2019-08 &  WFCCD \\
135210414 &  20:20:27.08 &  -36:34:25.79 &  17.214  &    0.086 &    AGN    & Y & N &  2019-09 &  NTT \\
135463347 &  20:13:23.96 &  -33:47:50.43 &  16.767  &    2.745 &    QSO    & Y & Y &  2019-09 &  NTT \\
135639174 &  20:46:44.37 &  -33:36:06.31 &  17.725  &    0.132 &    AGN    & Y & N &  2019-09 &  NTT \\
135758989 &  20:51:59.18 &  -32:28:38.67 &  17.577  &    1.974 &    QSO    & Y & Y &  2019-09 &  NTT \\
136204383 &  20:16:19.21 &  -27:33:21.66 &  17.458  &    2.062 &    QSO    & Y & N &  2019-09 &  NTT \\
136230878 &  20:19:46.01 &  -27:40:16.20 &  17.653  &    0.025 &    AGN    & Y & Y &  2019-09 &  NTT \\
136588662$^{b}$ &  20:11:58.77 &  -26:23:40.86 &  17.662  &    3.657 &    QSO    & Y & Y &  2019-09 &  NTT \\
136973888 &  20:55:54.12 &  -24:56:25.25 &  17.384  &    0.126 &    AGN    & Y & N &  2019-09 &  NTT \\
137044597 &  20:39:21.61 &  -24:00:05.00 &  17.942  &    2.916 &    QSO    & Y & Y &  2019-09 &  NTT \\
137166249$^{b}$ &  20:48:48.28 &  -22:51:52.07 &  17.449  &    3.100 &    QSO    & Y & Y &  2019-08 &  WFCCD \\
137203486 &  20:49:49.03 &  -21:46:33.63 &  17.299  &    0.038 &    AGN    & Y & N &  2019-09 &  NTT \\
171528002 &  20:19:31.38 &  -16:34:40.04 &  17.328  &    2.619 &    QSO    & Y & Y &  2019-09 &  NTT \\
172129292 &  20:40:43.96 &  -15:26:26.78 &  17.666  &    0.119 &    AGN    & Y & N &  2019-09 &  NTT \\
175467261 &  20:44:24.99 &  -01:48:55.56 &  17.335  &    2.304 &    QSO    & Y & Y &  2019-09 &  NTT \\
288304909 &  06:56:21.64 &  -74:21:02.00 &  17.359  &    2.116 &    QSO    & Y & Y &  2019-12 &  WFCCD \\
288408065 &  06:53:54.47 &  -75:07:56.30 &  17.339  &    3.081 &    QSO    & Y & Y &  2020-01 &  NTT/LDSS-3 \\
288742885 &  06:45:57.00 &  -67:53:09.09 &  17.720  &    4.314 &    QSO    & Y & Y &  2020-01 &  NTT \\
293711383$^{*}$ &  06:20:07.34 &  -61:55:57.93 &  18.021  &    3.320 &    QSO    & - & - &  2020-01 &  NTT \\
293925557 &  06:12:19.16 &  -55:44:21.29 &  17.298  &    2.285 &    QSO    & Y & Y &  2019-12 &  WFCCD \\
293946915$^{*}$ &  06:18:13.16 &  -55:19:46.12 &  18.049  &    3.117 &    QSO    & - & - &  2020-01 &  NTT \\
293970104$^{*}$ &  06:06:20.93 &  -54:19:38.59 &  18.210  &    3.343 &    QSO    & - & - &  2020-01 &  NTT \\
294463461 &  06:18:46.36 &  -51:08:16.66 &  17.173  &    2.050 &    QSO    & Y & Y &  2019-12 &  WFCCD \\
300818220 &  19:18:53.50 &  -52:50:11.91 &  17.311  &    2.022 &    QSO    & Y & Y &  2019-08 &  WFCCD \\
301296617 &  20:07:39.92 &  -58:10:37.65 &  17.744  &    3.289 &    QSO    & Y & Y &  2019-09 &  NTT \\
301508731 &  20:18:27.87 &  -51:51:00.35 &  17.060  &    2.639 &    QSO    & Y & Y &  2019-09 &  NTT \\
301643566 &  20:24:13.87 &  -53:12:13.87 &  16.962  &    2.208 &    QSO    & Y & Y &  2019-09 &  NTT \\
301677762 &  20:35:33.65 &  -52:53:29.14 &  17.746  &    2.572 &    QSO    & Y & Y &  2019-09 &  NTT \\
302676587 &  18:59:46.16 &  -74:34:32.82 &  17.310  &    0.142 &    AGN    & Y & N &  2019-09 &  NTT \\
303302739 &  20:12:29.32 &  -67:32:16.66 &  17.056  &    2.512 &    QSO    & Y & N &  2019-09 &  NTT \\
303604203 &  18:55:09.84 &  -81:38:28.84 &  17.456  &    1.746 &    QSO    & Y & N &  2019-09 &  NTT \\
304156429 &  20:25:23.01 &  -73:01:18.98 &  16.695  &    0.097 &    AGN    & Y & N &  2019-08 &  WFCCD \\
304169254 &  20:35:30.89 &  -75:38:11.99 &  17.705  &    2.195 &    QSO    & Y & Y &  2019-09 &  NTT \\
304191239 &  20:36:26.11 &  -78:16:34.13 &  17.031  &    2.627 &    QSO    & Y & Y &  2019-09 &  NTT \\
304867318 &  20:35:17.07 &  -62:37:01.82 &  16.730  &    2.709 &    QSO    & Y & N &  2019-09 &  NTT \\
305094211 &  21:10:38.23 &  -66:38:43.85 &  17.369  &    0.124 &    AGN    & Y & Y &  2019-09 &  NTT \\
305139405 &  21:32:46.07 &  -65:20:42.57 &  17.233  &    2.206 &    QSO    & Y & Y &  2019-08 &  WFCCD \\
305149856 &  20:36:41.07 &  -58:37:01.26 &  17.620  &    1.701 &    QSO    & Y & Y &  2019-09 &  NTT \\
305336573 &  21:08:17.67 &  -62:17:57.53 &  17.589  &    3.794 &    QSO    & Y & Y &  2019-09 &  NTT \\
305446236 &  21:40:40.86 &  -58:51:21.57 &  17.325  &    3.021 &    QSO    & Y & Y &  2019-09 &  NTT \\
305565087 &  21:57:46.57 &  -72:31:14.44 &  17.329  &    2.908 &    QSO    & Y & Y &  2019-09 &  NTT \\
305588367 &  21:58:49.70 &  -70:35:52.25 &  17.121  &    0.218 &    AGN    & Y & N &  2019-08 &  WFCCD \\
305669171 &  22:20:57.18 &  -69:59:23.17 &  17.432  &    2.221 &    QSO    & Y & Y &  2019-09 &  NTT \\
305744092 &  22:54:00.77 &  -72:07:48.71 &  17.584  &    2.922 &    QSO    & Y & Y &  2019-09 &  NTT \\
305772027 &  23:21:14.00 &  -73:26:02.07 &  17.103  &    0.143 &    AGN    & Y & N &  2019-08 &  WFCCD \\
305790724 &  22:34:32.52 &  -71:22:42.90 &  17.319  &    1.932 &    QSO    & Y & N &  2019-12 &  WFCCD \\
305817005 &  22:53:01.59 &  -69:38:13.41 &  17.725  &    3.082 &    QSO    & Y & Y &  2019-09 &  NTT \\
306048062 &  22:58:05.07 &  -65:59:07.40 &  17.712  &    2.796 &    QSO    & Y & Y &  2019-09 &  NTT \\
306142589 &  23:56:05.44 &  -66:09:49.56 &  17.622  &    3.143 &    QSO    & Y & Y &  2019-09 &  NTT \\
306255584 &  20:02:54.69 &  -42:14:21.06 &  17.516  &    2.612 &    QSO    & Y & Y &  2019-09 &  NTT \\
306376125${**}$ &  20:18:47.29 &  -45:46:48.42 &  16.462  &    1.355 &    QSO    & Y & N &  2019-06 &  LDSS-3/FIRE \\
306578934 &  20:40:01.98 &  -44:36:33.59 &  17.571  &    2.691 &    QSO    & Y & Y &  2019-09 &  NTT \\
306624073 &  20:46:36.95 &  -43:11:40.92 &  16.303  &    2.169 &    QSO    & Y & Y &  2019-09 &  NTT \\
306630517 &  20:46:33.08 &  -42:39:02.11 &  17.692  &    3.260 &    QSO    & Y & Y &  2019-09 &  NTT \\
306874313 &  21:01:47.82 &  -48:46:23.52 &  17.527  &    2.601 &    QSO    & Y & Y &  2019-09 &  NTT \\
307049678 &  21:18:23.60 &  -49:21:16.77 &  17.735  &    1.214 &    QSO    & Y & N &  2019-09 &  NTT \\
307108556 &  21:37:31.37 &  -47:38:03.44 &  16.769  &    2.448 &    QSO    & Y & Y &  2019-09 &  NTT \\
307443251 &  21:25:18.39 &  -42:05:47.62 &  17.363  &    3.549 &    QSO    & Y & Y &  2019-09 &  NTT \\
307536920 &  21:51:37.44 &  -44:36:44.17 &  17.363  &    3.638 &    QSO    & Y & Y &  2019-09 &  WFCCD \\
307772847 &  21:20:11.80 &  -53:04:07.41 &  17.556  &    2.245 &    QSO    & Y & Y &  2019-09 &  NTT \\
308002829 &  21:41:07.72 &  -52:30:39.28 &  17.287  &    2.334 &    QSO    & Y & Y &  2019-09 &  NTT \\
308038519 &  21:49:56.64 &  -50:31:28.72 &  17.743  &    3.296 &    QSO    & Y & Y &  2019-09 &  NTT \\
308280154 &  23:03:18.38 &  -59:15:33.20 &  17.554  &    1.408 &    QSO    & Y & N &  2019-12 &  LDSS-3 \\
308304837 &  23:17:51.93 &  -61:45:57.30 &  17.136  &    2.940 &    QSO    & Y & Y &  2019-08 &  WFCCD \\
308375290$^{a}$ &  23:35:05.86 &  -59:01:03.33 &  17.599  &    4.540 &    QSO    & Y & Y &  2019-09 &  NTT \\
308396429 &  23:44:05.92 &  -57:57:14.73 &  17.609  &    3.177 &    QSO    & Y & Y &  2019-09 &  NTT \\
308690703 &  22:41:31.39 &  -49:40:56.67 &  17.672  &    2.764 &    QSO    & Y & Y &  2019-09 &  NTT \\
308973148 &  23:44:57.76 &  -50:14:42.87 &  17.433  &    2.343 &    QSO    & Y & Y &  2019-12 &  LDSS-3 \\
308998359 &  22:58:55.77 &  -46:41:44.62 &  17.660  &    1.899 &    QSO    & Y & N &  2019-11 &  LDSS-3 \\
309030850 &  23:04:14.36 &  -44:23:17.76 &  17.679  &    2.283 &    QSO    & Y & Y &  2019-09 &  NTT \\
309431269 &  05:01:14.50 &  -80:19:03.28 &  17.936  &    2.663 &    QSO    & Y & Y &  2020-01 &  NTT \\
309455748 &  05:35:46.75 &  -78:15:08.61 &  17.685  &    2.939 &    QSO    & Y & Y &  2019-11 &  LDSS-3 \\
309656787 &  01:08:40.41 &  -82:17:47.76 &  17.579  &    2.707 &    QSO    & Y & Y &  2019-09 &  NTT \\
309939096 &  03:54:32.05 &  -74:03:44.86 &  17.510  &    0.000 &    STAR   & Y & N &  2019-09 &  NTT \\
309972073 &  02:18:10.04 &  -75:39:43.01 &  17.699  &    0.000 &    STAR   & Y & N &  2019-09 &  NTT \\
310047722 &  02:50:56.99 &  -72:04:58.62 &  17.602  &    2.892 &    QSO    & Y & Y &  2019-09 &  NTT \\
310206031 &  05:09:43.13 &  -74:09:47.89 &  17.575  &    3.773 &    QSO    & Y & Y &  2020-01 &  NTT \\
311228175 &  04:31:45.85 &  -74:51:26.83 &  17.189  &    3.140 &    QSO    & Y & Y &  2019-12 &  WFCCD \\
312716357 &  05:34:58.75 &  -64:49:36.83 &  17.541  &    3.247 &    QSO    & Y & Y &  2020-01 &  NTT \\
313912389 &  04:39:23.74 &  -65:36:22.83 &  17.805  &    3.399 &    QSO    & Y & Y &  2020-01 &  NTT \\
313928581 &  04:20:32.51 &  -66:09:30.84 &  17.260  &    2.700 &    QSO    & Y & Y &  2019-12 &  WFCCD \\
314899504 &  02:24:11.35 &  -70:36:38.67 &  17.739  &    2.829 &    QSO    & Y & Y &  2019-09 &  NTT \\
315156166 &  01:24:25.20 &  -66:08:13.38 &  17.554  &    2.365 &    QSO    & Y & N &  2019-12 &  LDSS-3 \\
315316671 &  02:43:04.74 &  -66:23:11.69 &  17.649  &    2.055 &    QSO    & Y & Y &  2019-09 &  NTT \\
315435952 &  03:12:32.49 &  -62:38:59.70 &  16.845  &    3.016 &    QSO    & Y & Y &  2019-09 &  NTT \\
315483506 &  02:24:56.71 &  -63:31:04.98 &  17.716  &    2.706 &    QSO    & Y & Y &  2019-09 &  NTT \\
315552288 &  02:37:48.15 &  -59:59:51.14 &  17.695  &    2.756 &    QSO    & Y & Y &  2019-09 &  NTT \\
315607762 &  03:17:24.89 &  -57:36:19.01 &  17.922  &    3.844 &    QSO    & Y & Y &  2020-01 &  NTT \\
315783093 &  05:47:55.85 &  -58:40:37.81 &  17.430  &    0.000 &    STAR   & N & N &  2019-12 &  WFCCD \\
315809832 &  05:27:04.37 &  -61:54:57.45 &  16.949  &    1.879 &    QSO    & Y & Y &  2019-12 &  WFCCD \\
315974502 &  05:34:49.40 &  -57:47:26.92 &  17.085  &    2.880 &    QSO    & Y & Y &  2019-12 &  WFCCD \\
316155633 &  05:08:24.66 &  -55:44:40.37 &  17.979  &    2.466 &    QSO    & Y & Y &  2020-01 &  NTT \\
316186672 &  05:01:00.45 &  -54:26:49.97 &  17.093  &    2.600 &    QSO    & Y & Y &  2019-12 &  WFCCD \\
316292063 &  05:48:03.20 &  -48:48:13.19 &  16.886  &    4.147 &    QSO    & Y & Y &  2019-12 &  WFCCD \\
316293416 &  05:45:02.05 &  -49:24:47.35 &  17.344  &    2.928 &    QSO    & Y & Y &  2020-02 &  LDSS-3 \\
316383429 &  05:55:08.16 &  -45:58:23.61 &  17.649  &    3.262 &    QSO    & Y & Y &  2020-01 &  NTT \\
316434053 &  05:39:21.91 &  -45:16:58.03 &  17.064  &    2.556 &    QSO    & Y & Y &  2019-12 &  WFCCD \\
316563116$^{*}$ &  05:20:23.95 &  -47:52:12.06 &  18.362  &    3.825 &    QSO    & - & - &  2020-01 &  NTT \\
316567405$^{*}$ &  05:24:10.76 &  -47:15:28.25 &  18.052  &    2.399 &    QSO    & - & - &  2020-01 &  NTT \\
316591563 &  05:29:14.28 &  -45:08:07.03 &  17.661  &    3.690 &    QSO    & Y & Y &  2020-01 &  NTT \\
316663640 &  04:30:07.62 &  -56:47:36.38 &  17.154  &    3.130 &    QSO    & Y & Y &  2019-12 &  WFCCD \\
316757721 &  04:33:26.90 &  -52:07:58.20 &  17.777  &    3.442 &    QSO    & Y & Y &  2020-01 &  NTT \\
316769552 &  04:18:15.82 &  -52:58:14.21 &  15.171  &    0.000 &    STAR   & N & N &  2019-12 &  WFCCD \\
316788893 &  04:25:40.68 &  -50:43:15.05 &  17.528  &    0.047 &    AGN    & Y & Y &  2019-11 &  WFCCD \\
316791881 &  04:11:46.66 &  -52:12:07.74 &  17.605  &    2.022 &    QSO    & Y & Y &  2019-11 &  LDSS-3 \\
316933237 &  03:45:05.36 &  -49:35:14.61 &  17.656  &    0.000 &    STAR   & Y & N &  2019-09 &  NTT \\
317020875 &  04:45:18.99 &  -46:00:49.84 &  17.423  &    2.184 &    QSO    & Y & Y &  2019-12 &  WFCCD \\
317068530 &  04:14:59.70 &  -49:07:56.59 &  17.652  &    2.790 &    QSO    & Y & Y &  2019-11 &  LDSS-3 \\
317105305 &  03:56:39.60 &  -47:08:34.05 &  17.407  &    2.110 &    QSO    & Y & Y &  2019-12 &  WFCCD \\
317138695 &  04:30:48.53 &  -43:28:35.18 &  17.660  &    2.757 &    QSO    & Y & Y &  2020-01 &  NTT \\
317201391 &  00:42:44.49 &  -62:41:09.29 &  17.464  &    2.055 &    QSO    & Y & N &  2019-12 &  LDSS-3 \\
317253125 &  00:48:05.34 &  -59:29:09.44 &  17.536  &    3.607 &    QSO    & Y & Y &  2019-09 &  NTT \\
317343050 &  01:27:16.87 &  -58:02:47.28 &  17.772  &    3.918 &    QSO    & Y & Y &  2020-01 &  NTT \\
317358346$^{a}$ &  00:07:36.56 &  -57:01:51.73 &  17.240  &    4.260 &    QSO    & Y & Y &  2019-08 &  WFCCD \\
317411112 &  00:18:30.46 &  -53:35:35.20 &  17.744  &    3.738 &    QSO    & Y & Y &  2019-09 &  NTT \\
317506585 &  01:42:17.55 &  -59:37:00.86 &  17.717  &    3.432 &    QSO    & Y & Y &  2019-09 &  NTT \\
317647507 &  01:40:04.35 &  -54:23:15.53 &  17.623  &    2.458 &    QSO    & Y & Y &  2019-09 &  NTT \\
317802750$^{a}$ &  00:12:24.99 &  -48:48:29.86 &  17.715  &    4.621 &    QSO    & Y & Y &  2019-09 &  NTT \\
318357314 &  03:45:42.68 &  -47:36:31.41 &  17.590  &    2.878 &    QSO    & Y & Y &  2019-09 &  NTT \\
318395533 &  03:45:07.90 &  -44:32:44.08 &  17.192  &    2.140 &    QSO    & Y & N &  2019-12 &  WFCCD \\
318397429 &  03:43:30.07 &  -43:44:29.86 &  17.121  &    1.777 &    QSO    & Y & Y &  2019-12 &  WFCCD \\
318432798 &  03:54:35.83 &  -42:16:30.70 &  16.737  &    1.983 &    QSO    & Y & N &  2019-12 &  WFCCD \\
318446298 &  03:23:24.22 &  -44:03:46.17 &  17.473  &    2.074 &    QSO    & Y & Y &  2020-01 &  NTT \\
318479392 &  03:18:24.60 &  -41:32:40.30 &  17.647  &    2.495 &    QSO    & Y & Y &  2019-11 &  WFCCD \\
318495154 &  03:38:35.49 &  -40:30:41.39 &  17.309  &    2.900 &    QSO    & Y & Y &  2019-12 &  WFCCD \\
318699655 &  03:21:50.82 &  -39:53:14.75 &  17.471  &    2.794 &    QSO    & Y & Y &  2019-12 &  WFCCD \\
\enddata
\tablecomments{
      \small
      \\
      $^{a}$Independently discovered also by \cite{Wolf19} \\
      $^{b}$Included also in \cite{Sch19b} \\
      $^{c}$Included also in \cite{Sch19a} \\
      $^{*}$ Not part of the Main Sample \\
      $^{**}$ after updated reduction of PaperI observed sources \\ 
      }
\end{deluxetable*}
\end{longrotatetable}

\end{document}